%% file: main.tex
\documentclass[sigconf]{acmart}

\AtBeginDocument{%
  \providecommand\BibTeX{{%
    \normalfont B\kern-0.5em{\scshape i\kern-0.25em b}\kern-0.8em\TeX}}}

\usepackage{graphics}
\usepackage{stfloats}
\usepackage{url}
\copyrightyear{2025}
\acmYear{2025}
\setcopyright{cc}
\setcctype{by}
\acmConference[CHI '25]{CHI Conference on Human Factors in Computing Systems}{April 26-May 1, 2025}{Yokohama, Japan}
\acmBooktitle{CHI Conference on Human Factors in Computing Systems (CHI '25), April 26-May 1, 2025, Yokohama, Japan}\acmDOI{10.1145/3706598.3714181}
\acmISBN{979-8-4007-1394-1/25/04}

\newcommand \change[1]{{\textcolor{black}{#1}}}
\newcommand \del[1]{}

\newcommand{\name}{EyeLingo}

\begin{document}

\title[Unknown Word Detection for English as a Second Language (ESL) Learners ...]{Unknown Word Detection for English as a Second Language (ESL) Learners Using Gaze and Pre-trained Language Models}


\author{Jiexin Ding}
\email{jxding17@gmail.com}
\affiliation{%
  \department{Department of Computer Science and Technology, Global Innovation Exchange (GIX) Institute}
  \institution{Tsinghua University}
  \city{Beijing}
  \country{China}
}
\affiliation{%
  \department{Electrical \& Computer Engineering}
  \institution{University of Washington}
  \city{Seattle}
  \state{WA}
  \country{USA}
}

\author{Bowen Zhao}
\email{bowen@groundlight.ai}
\affiliation{%
  \institution{Groundlight AI}
  \city{Seattle}
  \state{WA}
  \country{USA}
}

\author{Yuntao Wang}
\email{yuntaowang@tsinghua.edu.cn}
\affiliation{%
  \department{Department of Computer Science and Technology}
  \institution{Tsinghua University}
  \city{Beijing}
  \country{China}
}
\authornote{denotes as the corresponding author.}

\author{Xinyun Liu}
\email{liuxinyun6@yahoo.com}
\affiliation{%
  \institution{Rice University}
  \city{Houston}
  \state{TX}
  \country{USA}
}

\author{Rui Hao}
\email{haorui24@mails.ucas.ac.cn}
\affiliation{%
  \department{School of Artificial Intelligence}
  \institution{University of Chinese Academy of Sciences}
  \city{Beijing}
  \country{China}
}

\author{Ishan Chatterjee}
\email{ichat@cs.washington.edu}
\affiliation{%
    \department{Paul G. Allen School of Computer Science and Engineering}
   \institution{University of Washington}
  \city{Seattle}
  \state{WA}
  \country{USA}
}

\author{Yuanchun Shi}
\email{shiyc@tsinghua.edu.cn}
\affiliation{%
  \department{Department of Computer Science and Technology}
  \institution{Tsinghua University}
  \city{Beijing}
  \country{China}
}
\affiliation{%
  \institution{Qinghai University}
  \city{Xining}
  \country{China}
}

\renewcommand{\shortauthors}{XXX, et al.}


\begin{abstract}
English as a Second Language (ESL) learners often encounter unknown words that hinder their text comprehension. Automatically detecting these words as users read can enable computing systems to provide just-in-time definitions, synonyms, or contextual explanations, thereby helping users learn vocabulary in a natural and seamless manner.
This paper presents \name, a transformer-based machine learning method that predicts the probability of unknown words based on text content and eye gaze trajectory in real time with high accuracy. 
A 20-participant user study revealed that our method can achieve an accuracy of 97.6\%, and an F1-score of 71.1\%.
We implemented a real-time reading assistance prototype to show the effectiveness of \name. The user study shows improvement in willingness to use and usefulness compared to baseline methods.
\end{abstract}

\ccsdesc[500]{Human-centered computing~Interaction techniques}

\keywords{Unknown word detection, gaze, pre-trained language model.}


\maketitle

\input{sections/1-intro}

\input{sections/2-related_work}

\input{sections/3-method}

\input{sections/4-experiment}
\input{sections/5-evaluation}
\input{sections/7-discussion}

\input{sections/8-conclusion}

\begin{acks}
This work is supported by the Natural Science Foundation of China under Grant No. 62132010, 62472244, 62102221, the Tsinghua University Initiative Scientific Research Program, and the Undergraduate Education Innovation Grants, Tsinghua University.
\end{acks}

\bibliographystyle{ACM-Reference-Format}
\bibliography{ref}

\end{document}

%% file: sections/1-intro.tex
\section{Introduction}

\begin{figure*}
  \centering
  \includegraphics[width = 0.9\linewidth]{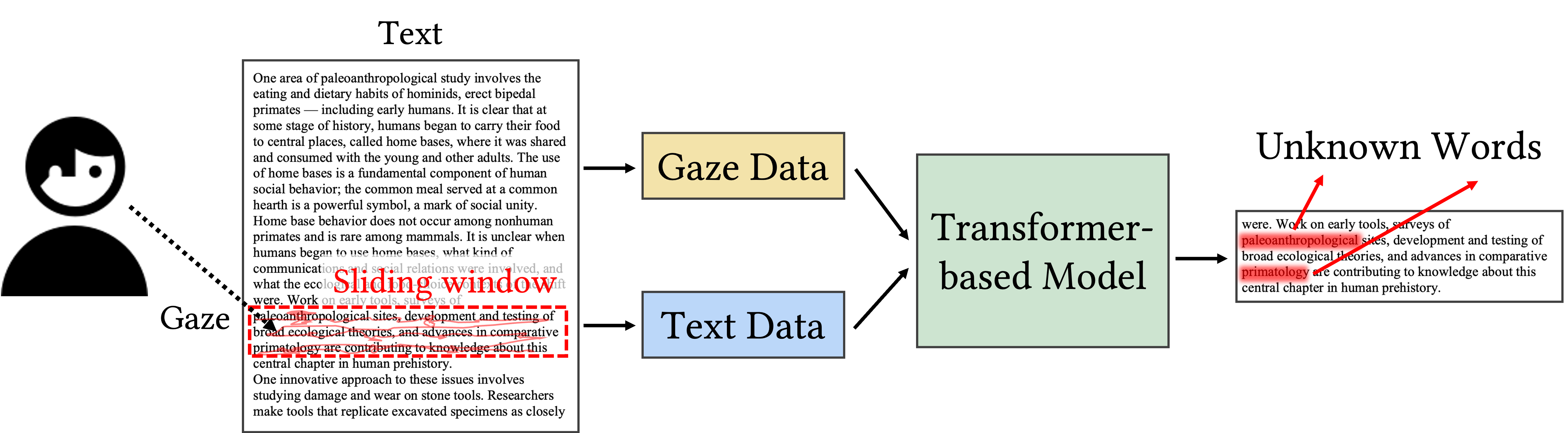}
  \caption{Our method locates the content the user is reading in real-time through gaze, and inputs the gaze data and text data to the transform-based model to detect unknown words.}
  \label{fig:intro} 
\end{figure*}

Unknown words can greatly reduce reading fluency and worsen the reading experience of non-native speakers~\cite{rigg1991whole,mokhtar2012guessing}. By automatically detecting these unknown words as users read, computing systems can assist users in reading and language comprehension, and provide just-in-time word explanations for learning vocabulary. Because unknown words vary among users, many previous works were based on the explicit expression of intention by users, such as mouse clicks~\cite{web_ehara_2010} or looking at words intentionally~\cite{dwell_time, eye_tracking_dussias}. To facilitate a more natural reading experience, other methods use gaze characteristics such as fixation duration, number of fixations, and saccade length~\cite{sibert_reading_2000, Reading_Assistanc_Guo} to detect unknown words automatically, since there is a correlation between gaze and word difficulty~\cite{just1980theory}. Previous research shows that fixation, the maintaining of the gaze on a single location, happens when people are processing phrases~\cite{just1980theory}. However, these gaze-based methods have two major challenges affecting their accuracy and ease of deployment. First, these methods require dedicated and costly eye-tracking hardware to obtain accurate eye movement data for these features. Moreover, even with professional eye trackers, measuring gaze is inherently inaccurate due to complicated eye motions, making it hard to precisely map a gaze point to the word in the text being read~\cite{eye_mice_bates_2003}. To reduce reliance on gaze information, other works seek to compensate or replace inaccurate eye-tracking data on commercial devices by incorporating text~\cite{gaze-text_garain_2017, unknown-word_hiraoka_2016, idict2006hyrskykari, du_using_2024}, click~\cite{web_ehara_2010} and motion data~\cite{imu_higa_2022}. 

Besides the cognitive process reflected by gaze, inherent linguistic information about words is also crucial for identifying unknown words. With the development of Natural Language Processing (NLP) technology, pre-trained language models (PLMs) demonstrate a powerful ability to capture rich linguistic information~\cite{devlin2019bert, liu2019roberta} which is strongly related to word difficulty~\cite{More_than_frequency}.
We explore how to take advantage of the language model in addition to gaze to make unknown word detection accurate, easily accessible, and more applicable. We present a real-time unknown word detection method that locates a region of interest based on gaze and then integrates linguistic information provided by PLM and gaze trajectory to predict unknown words using a transformer-based model. In this way the inaccuracy of the gaze-based method can be compensated by the probabilities distributed on words derived from the language model.

As shown in Fig.~\ref{fig:intro}, we tackle the problem of unknown word detection with two modalities of information, gaze and text. To process gaze patterns, we utilize the cutting-edge transformer-based encoder-decoder model with cross-attention modules to learn the positional information based on gaze trajectory and text layout. For linguistic information, we apply RoBERTa to the text in the region of interest with several crucial word-level knowledge embeddings (term frequency, part-of-speech, etc.). By jointly training the models above on our collected dataset, our approach surpasses existing methods with 71.1\% F1-scores and 97.6\% accuracy for unknown word detection.
With our method, real-time language learning assistance and just-in-time vocabulary acquisition tools can be enabled.

In this paper, we first verified the feasibility of our approach on the dataset collected by a professional eye tracker. We also applied our unknown word detection method to the relatively noisy webcam-based gaze data to show the robustness of our method with commodity deployments. Then, we conducted experiments to analyze the contribution of gaze and PLM. Finally, we implemented a reading assistance prototype to evaluate our method in a real world setting where data is processed in real time and investigate its usefulness and users' subjective experience.

Contributions in this work are summarized as follows:
\begin{enumerate}
    \item We propose an unknown word detection method (\name{}) that leverages gaze to locate a region of interest and then classifies wrods based on linguistic characteristics provided by a pre-trained language model and gaze encoding derived from a transformer-based model. It achieves an accuracy of 97.6\% and an F1-score of 71.1\%.
    \item We analyze the contributions of gaze and the pre-trained language model. Unknown word detection is mainly based on linguistic characteristics provided by a pre-trained language model. Gaze provides user-dependent and timely information to detect different unknown words for different users in a real-time manner.
    \item We build a reading assistance prototype and conduct a real-time evaluation. The result shows that the F1-score is 56.5\%. The user study shows improvement in willingness to use and usefulness compared to the traditional method.
\end{enumerate}

%% file: sections/2-related_work.tex
\section{Related Work}
In this section, we illustrate how gaze behavior is related to cognitive processes in reading and examine the reading behavior analysis enabled by gaze tracking. Then, we explain the limitations of the gaze-based unknown word detection method and how the previous works improved their performance. After that, we analyze the gaps in detecting unknown words accurately.

\subsection{Gaze in Reading}
\label{sec:related_word_gaze}
Reading as a cognitive process affects eye movement~\cite{just1980theory}. When people are reading, the eyes follow the text through small amplitudes, ballistic motions called saccades. The pauses between two saccades are called a fixation~\cite{idict2006hyrskykari}. According to the eye-mind hypothesis, the fixation on a word persists during its processing phase~\cite{just1980theory}. Consequently, fixation duration can be a metric for identifying difficult words and measuring cognitive processes in reading. For this reason, previous research combines fixation duration with other gaze features to predict reading comprehension~\cite{cheng_gaze-based_2015, okoso_towards_2015,sanches_using_2017}, detecting mind wandering~\cite{bixler_automatic_2016}, identifying interest~\cite{wikigaze_2020_dubey}, and detecting attention~\cite{li_multimodal_2016, zermiani-etal-2024-interead, hollenstein-etal-2020-zuco} in reading. Other works facilitate natural language processing tasks such as named entity recognition and sequence classification by integrating gaze into neural network~\cite{sentiment_long_21,barrett2020sequence, barrett-etal-2018-sequence, ner_Hollenstein_2019}. Besides, some researches also improve the prediction of gaze behavior such as scanpath and calibration process based on the connection between gaze and text~\cite{TSM_NEURIPS2020_460191c7,eyettention_2023, CalibRead_2024_liu}. However, most of these works focus on paragraph-level and sentence-level tasks rather than word detection.

Extended periods of fixation on the focal word suggest difficulties in word identification~\cite{idict2006hyrskykari}, forming the theoretical basis for the detection of unknown words via gaze. However, it is hard to achieve high accuracy only based on the fixation because of the inaccuracy of gaze-tracking hardware, algorithms, and the ambiguous relationship between gaze patterns and the cognitive processes of understanding words. The highest accuracy of eye tracking is around $0.3^\circ$ (2.6 mm when the distance is 50 cm) under optimal conditions, but the accuracy can easily be affected by the calibration performance and user posture~\cite{eye_tracking2022liu}. Thus, how to accurately match the gaze point to the focused text is a problem. Additionally, other researchers pointed out that the processing of words can occur when they are not held in fixation~\cite{rayner1998eye}. 

Although there is a strong correlation between gaze behavior and word difficulty, the above limitations make it difficult to achieve high accuracy in detecting unknown words based only on gaze.
Previous gaze-based unknown word detection methods improve their accuracy by combining multiple gaze features and leveraging text information such as word length and word rarity~\cite{unknown-word_hiraoka_2016, gaze-text_garain_2017}. 

\subsection{Unknown Word Detection}

According to the importance of gaze in analyzing reading behavior as explained above, most of the related works are gaze-based. iDict~\cite{idict2006hyrskykari} detects unknown words based on gaze duration and word frequency and sets a threshold to trigger a gloss or margin note on unknown words. It successfully detects 36.5\% of unknown words. Later works extend iDict by replacing the threshold function with machine learning. Hiroka et al.~\cite{unknown-word_hiraoka_2016} uses several gaze features such as first gaze duration, number of fixations, and number of regressions and feeds them into support vector machines (SVM) to classify the unknown word. Their model performs best (F1-score of 55.6\%) when adding linguistics features including word length and word rarity. Similarly, Garain et al.~\cite{gaze-text_garain_2017} also take both gaze and linguistics features into consideration to achieve the best F1-score of 86\% on a single user using an SVM. However, each of these methods rely on a professional eye tracker to obtain sufficiently accurate gaze data to extract  gaze features and match these features to the words. Even with a dedicated eye tracker, it is still difficult to accurately assign each gaze point to the corresponding line due to the vertical error~\cite{yamaya2015dynamic, Sanches2016Vertical}. Therefore, in their experiment, the line spacing was set between 3.0 to 6.0 to better distinguish lines based on the y-value. The large line spacing makes it difficult for their method to be applied in real-world scenarios, considering that the line spacing of text we usually read is mostly between 1.0 and 2.0. Moreover, their data preprocessing that relies on global gaze coordination makes them even harder to be applied in a real-time application.

Apart from gaze, other reading behaviors such as mouse clicks and hand motion can be used to detect unknown words. Ehara et al.~\cite{web_ehara_2010} gives users potential highlights in advance and analyzes users' feedback based on their clicks on the web page. Predictions are improved based on this feedback, reaching an accuracy of up to 80.01\%. Higashimura et al.~\cite{imu_higa_2022} targets at vocabulary acquisition on smartphones and identifies unknown words utilizing the motion data obtained from the inertia sensors on smartphones. The estimation improves through the reading and the AUPR is about 0.3. Both methods are device-specific and only applicable to a single user as they need to be optimized based on personalized iterative feedback for higher accuracy.

In summary, the current best-performing detection methods combine gaze and text data. However, these current methods still rely on heavy gaze data preprocessing and special experimental settings. A real-time method that is robust to relatively inaccurate gaze data can promote the availability of unknown-word detection. Since word difficulty is highly dependent on linguistic features\cite{More_than_frequency}, we seek to build upon the rapid development of natural language processing (NLP) technology~\cite{attention_is_all_you_need} in recent years to yield a solution. Pre-trained language models such as ~\cite{devlin2019bert,liu2019roberta} encompass extensive text information for downstream applications. We harness their capabilities in a new architecture to increase the tolerance of inaccuracy in gaze tracking. We propose a transformer-based model that requires linguistic information provided by PLM from a region of interest around the target words and learns the gaze pattern automatically from gaze trajectory and text position.

%% file: sections/3-method.tex
\section{Unknown Word Detection Method}
\label{sec:method}

In this section, we explain why we chose gaze and PLM to detect unknown words. Then we describe how we collected and processed the data. Finally, we explain the architecture of our method in detail.

\subsection{Method Justification}
\begin{figure*}
  \centering
   \includegraphics[width = 0.72\linewidth]{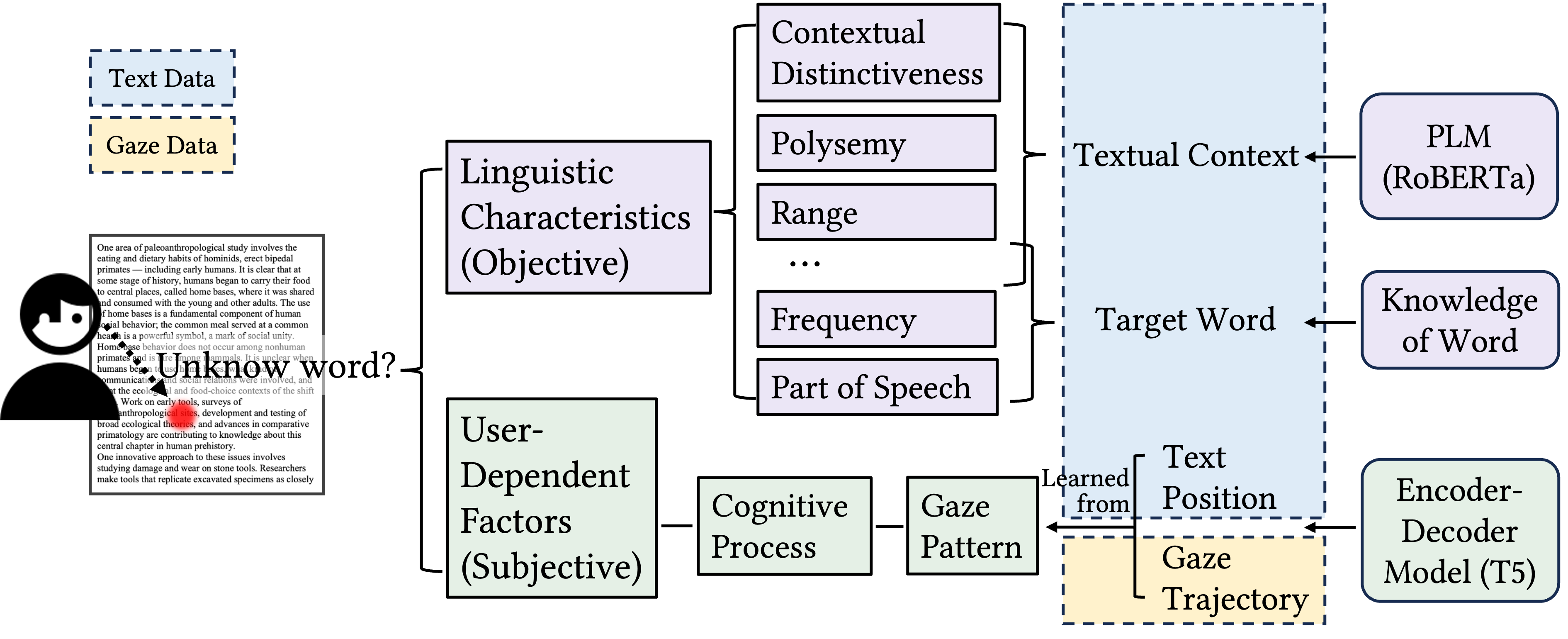}
  \caption{Methodology for Detecting Unknown Words Using Integrated Text and Gaze Information. This integrated approach leverages both linguistic and user-dependent factors to effectively identify unknown words.}
  \label{fig:justification} 
\end{figure*}

Even though whether a word is unknown to a user is subjective, linguistic priors related to word difficulty are helpful in unknown word prediction. Therefore, we assume a user's unknown word can be predicted based on (1) linguistic characteristics related to word difficulty, such as frequency, contextual distinctiveness, and COCA (Corpus of Contemporary American English) range, and (2) user-dependent factors which can be reflected by gaze trajectory. Therefore, we integrate text and gaze information to detect unknown words as shown in Fig.~\ref{fig:justification}:
\begin{enumerate}
    \item Linguistic characteristics of the target word's tokens and the textual context in the region of interest are captured by RoBERTa. RoBERTa is an effective and efficient language model for natural language understanding, ideally fitting our expectations, based on its proven performance in capturing complex linguistic features, such as contextual embeddings, syntactic structures, and semantic relationships. Prior research demonstrated that models like RoBERTa outperform traditional methods in tasks requiring a deep understanding of natural language~\cite{liu2019roberta}. Considering that the context around the word influences word difficulty~\cite{More_than_frequency}, we applied RoBERTa to all the text in the region of interest. Besides, as we aim at deploying \name{} to end-user devices like laptops, we do not consider using billion-level large language models for efficiency reasons. Meanwhile, generative models are also unsuitable in our use case, as our task is to identify users' unknown words in the given text rather than generating new content. Therefore, we choose RoBERTa instead of GPT. The model details related to this are explained in Section~\ref{sec:model_text}.
    \item Word-level knowledge of the target word such as term frequency, part of speech and named entity recognition are applied to enhance the performance. The tokenization is required to use RoBERTa. The word-level linguistic characteristics are partially lost during tokenization. Thus, we introduce word-level knowledge to compensate for the loss. The model details related to this are explained in Section~\ref{sec:model_knowledge}.
    \item Gaze pattern is automatically learned from gaze trajectory and text position by the model adapted from T5, one of the top encoder-decoder transformer-based architectures. Gaze patterns can reflect a user’s cognitive process, indicating whether the user encounters an unknown word~\cite{just1980theory}. Considering the complexity of gaze patterns and the variability between users, we want to automatically learn these patterns using machine learning models based on gaze trajectory and text position. The encoder-decoder model can learn complex relationships between different types of sequential data. T5 has demonstrated cutting-edge performance on a wide range of tasks and is capable of capturing complex patterns and dependencies within input sequences~\cite{raffel2020exploring}. We take out the output of the last encoder layer as the positional encoding which represents the relationship between the gaze trajectory and text position. The model details related to this are explained in Section~\ref{sec:model_positional_data_encoding}.
\end{enumerate}

Our method has three main advantages compared to previous works:
\begin{enumerate}
    \item \name{} has a higher tolerance for the noise of gaze data and has better performance by leveraging language models' ability to capture linguistic characteristics to reduce the reliance on gaze (results in Section~\ref{sec:main_result} and Section~\ref{sec:contribution_PLM}).
    \item \name{} eliminates tedious feature engineering by automatically learning gaze patterns based on gaze trajectory and text coordinates (detailed explanations in Section~\ref{sec:model_positional_data_encoding} and results in Section~\ref{sec:contribution_gaze}).
    \item \name{} enables real-time detection by only relying on the local information in the region of interest, which is benefit from the low reliance on gaze (results in Section~\ref{sec:user_evaluation}).
\end{enumerate}

\subsection{Unknown Word Detection Model}
\label{sec:method_model}
\begin{figure*}[htbp]
  \centering
  \includegraphics[width = 0.9\linewidth]{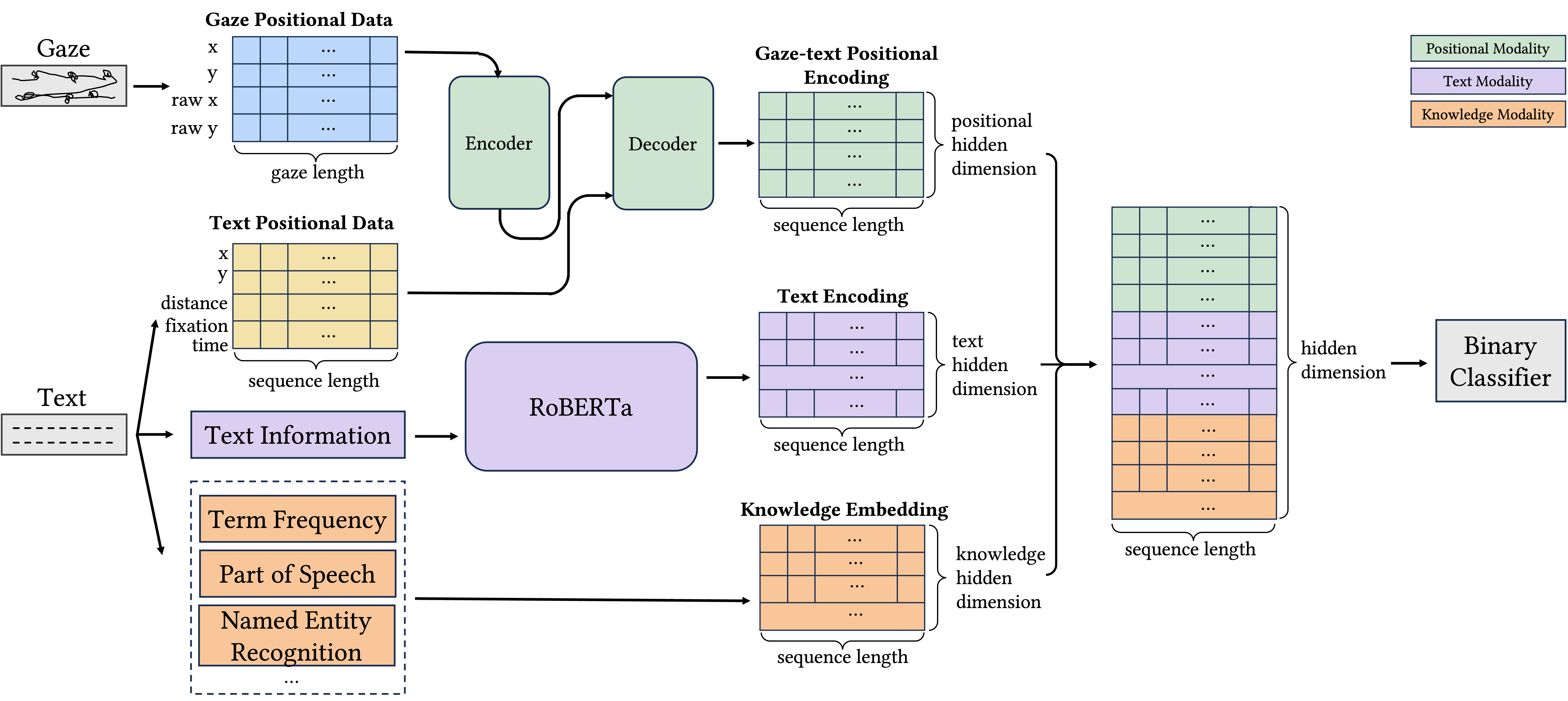}
  \caption{Our model includes an encoder-decoder model to encode positional data, a pre-trained RoBERTa to encode text information, and learnable embeddings to encode the knowledge. The concatenation of these three matrices is input to a binary classifier.}
  \label{fig:model} 
\end{figure*}

The goal of our model is to classify whether a word is an unknown word or not using both gaze information and text information. We use the encoder-decoder architecture to encode positional gaze and text information into the vector space. We also leverage RoBERTa~\cite{liu2019roberta} to integrate with text information. Moreover, we also add the prior knowledge to take the use of the word-level information. The overall architecture of our model is shown in Fig.~\ref{fig:model}.

\subsubsection{Positional Data Encoding}
\label{sec:model_positional_data_encoding}
To accurately capture the correlation between the user's gaze and the document, we adapt the model architecture from T5~\cite{raffel2020exploring}, the state-of-the-art encoder-decoder language model to process the positional data from those two modalities, where the encoder learns to capture the user's gaze pattern and the decoder is expected to predict whether a token belongs to an unknown word of the user, based on the encoder's outputs and the tokens' positional data. For the encoder part, we feed the raw and the moving averaged gaze trace to let the model capture both the fine-grained and general positional information from the user's behavior. As for the decoder part, instead of using the token-level positional data as the inputs only, we also calculate the averaged gaze-token distance $d(g, w)$ and the gaze duration $t(g, w)$ (the length of the time when the user's gaze lives in the bounding box of the token) for each token $w$
\begin{align}
    d(g, w) &= \sqrt{(\frac{\sum_{i=1}^{N_g} g_x^i}{{N_g}} - \frac{w_x^s + w_x^t}{2}) ^ 2 + (\frac{\sum_{i=1}^{N_g} g_y^i}{{N_g}} - \frac{w_y^s + w_y^t}{2}) ^ 2} \\
    t(g, w) &= |\{(g_x^i, g_y^i)| 1 \le i \le N_g \land w_x^s \le g_x^i \le w_x^t \land w_y^s \le g_y^i \le w_y^t\}|
\end{align}
where $N_g$ is the number of gaze samples within the sliding window, $w_x^s, w_x^t, w_y^s, w_y^t$ are the coordinates of the bounding box of the token $w$, and $g_x$ and $g_y$ are the user's gaze trace on x and y axis, respectively. Overall, this encoder-decoder model can be summarized as
\begin{align}
    H_g &= \text{Encoder}(g_x, g_y, g_x^{raw}, g_y^{raw}) \\
    P &= \text{Decoder}(H_g, w_x, w_y, d(g, w), t(g, w))
\end{align}
where $g_x, g_y$ are the moving-averaged gaze positional data, $g_x^{raw}, g_y^{raw}$ are the raw gaze data without filters applied, $w_x, w_y$ are the positions of each token, and $H_g$ is the final encoder outputs, which is used as the inputs (more specifically, keys and values) of the cross-attention modules in decoder layers.

\subsubsection{Textual Information Capturing}
\label{sec:model_text}
We utilize a pre-trained RoBERTa, a widely used pre-trained language model based on the transformer architecture, to encode the text information. The layer consists of a self-attention module and a feed-forward layer. The structure can help the model better encode the text by using the surrounding text to establish the context. It encodes text data $s \in \mathbb{R}^{n_{txt}}$ to $Z \in \mathbb{R}^{n_{txt} \times n_r}$, in which $n_r$ is the hidden dimension of RoBERTa.

\begin{equation}
Z = \textrm{RoBERTa}(s)
\end{equation}

\subsubsection{Knowledge-grounded Enhancement}
\label{sec:model_knowledge}
The pre-trained language model takes tokens instead of words as input. Word-level information may be lost during tokenization. Therefore, we introduce word-level knowledge including the term frequencies, part of speech~\cite{bird2009natural}, and named entity recognition~\cite{honnibal2020spacy} to utilize word-level information. These features are encoded into a knowledge matrix $K \in \mathbb{R}^{n_{txt} \times n_k}$.

\subsubsection{Training}
We combine the encodings of the three modules' outputs as the inputs for the final classifier layer, a logistic regression module for binary classification. We fully fine-tune our model, including the positional encoder-decoder module, the RoBERTa pre-trained module, the knowledge embeddings, along with the final classifier on our dataset. It shall be noticed that the task of unknown word detection is significantly class imbalanced, where on average there are 15 times more negative tokens (known words) than positive ones (unknown words). To mitigate this issue, we use the focal binary entropy loss for our model's training which re-weights the loss term for more robust training:
\begin{align}
    H &= [P; Z; K] \\
    p &= \sigma(W_o \cdot H+ b_o) \\
    \mathcal{L}(p, \hat{y}) &= - \alpha \hat{y} (1-p)^\gamma \log(p) - (1-\alpha) (1-\hat{y}) p^\gamma \log(1-p) 
\end{align}
where $p$ is the model's prediction logits, $\hat{y}$ is the ground truth, while $\alpha$ and $\gamma$ are the hyper-parameters that control the weight between the two classes and the speed of the model's focus on difficult examples, respectively.

\subsection{Data Preparation}
\label{sec:method_data_preparation}
\subsubsection{Implementation}
\label{sec:method_implementation}
We built a system to collect gaze data from eye trackers and webcams at the same time. We used Tobii Pro Nano\footnote{https://www.tobii.com/products/eye-trackers/screen-based/tobii-pro-nano} eye tracker whose sampling rate is 60Hz and streamed its data to the computer through a Python script. The accuracy of the optimal conditions of Tobii Nano eye tracker is $0.3^\circ$. For webcam data, we used a SeeSo\footnote{https://seeso.io/}, a remote eye tracking platform, by integrating it into our PDF reader. To get text information including their contents and positions, we built a web-based PDF reader based on an open source Github repository\footnote{https://github.com/zotero/pdf-reader}. This platform can record gaze data using Seeso, retrieve text information when users finish reading, and record users' clicks while they are labeling their unknown words after the reading. 

The laptops we used were Macbook Pro (CPU: Apple M1 Pro, RAM: $16$ GB, Storage: $512$ GB, Screen Size: $14.2$ inches, Resolution: $1512 \times 982$)\footnote{https://support.apple.com/en-us/111902} and Huawei MateBook D14 2022 (CPU: i5-1155G7, RAM: $16$ GB, Storage: $512$ GB, Screen Size: $14$ inches, Resolution: $1920 \times 1080$)\footnote{https://consumer.huawei.com/en/laptops/matebook-d-14-2022/specs/}. The height of the line is $20.63$ pixels on Macbook and $20.74$ on Matebook, so the height of each line of text is approximately $4.1$ mm on the MacBook and about $3.3$ mm on the MateBook during the experiment. According to the Tobii calibration requirement\footnote{\url{https://connect.tobii.com/s/article/how-to-position-participants-and-the-eye-tracker?language=en_US}}, the distance between the participant's face and the screen during the experiment is about $50-60$ cm. Thus, the 10-point font size can be converted to $0.39^\circ - 0.47^\circ$ on a MacBook and $0.32^\circ - 0.39^\circ$ on a MateBook.

\begin{figure}[htbp]
  \centering
  \includegraphics[width=1.0\columnwidth]{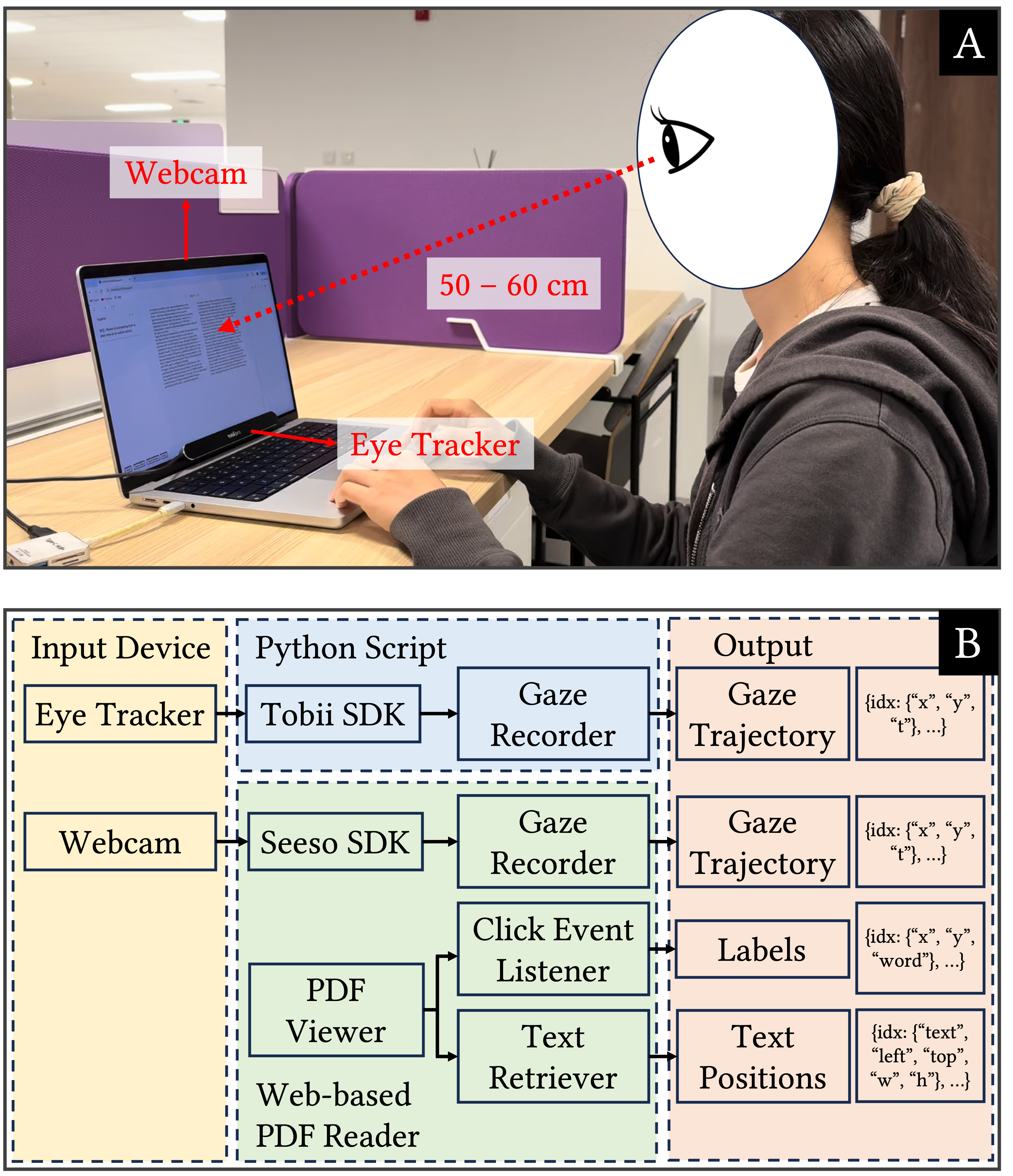}
  \caption{(A) During data collection, users are seated at a distance of approximately 50-60 cm from the screen, as instructed by the Tobii calibration interface. The eye tracker and webcam simultaneously collect data. (B) The data collection platform includes a Python script to read eye tracker data and a web-based PDF reader to read webcam data and record word labeling.}
  \label{fig:implementation} 
\end{figure}

\subsubsection{Participant and Material}
We recruited a total of 20 undergraduate and graduate students (5 females, 15 males) whose second language is English. Their Vocabulary Levels Test (VLT) scores ranged from 7 to 30 ($M = 20.24, SD = 7.15$). The VLT is widely used to measure the vocabulary level of English learners~\cite{nation1990teaching,schmitt2001developing}. We chose the 5000-word frequency level group because the TOEFL reading materials we used for the user study presume a vocabulary knowledge of approximately 4500 words~\cite{chujo2009many}. Their ages range from 21 to 26 years old (\textit{M} = 22.85, \textit{SD} = 1.65).
Among them, 16 people wore glasses during data collection and 4 did not.
The reading materials contain 120 articles of TOEFL and GRE reading with an average length of 363 words per article and 43534 words in total. We organized the text into a common paper format, which is single-spaced and has two columns. The font is Times New Roman and the font size is 10.

\subsubsection{Experiment Design and Procedure}
\label{sec:method_procedure}
To increase text diversity, we divided twenty participants into four groups, and participants in each group read the same 30 articles. These 30 articles were divided into 3 days to read, which took about an hour each day. There was a calibration session before data collection for each participant each day to calibrate both the eye tracker and webcam eye tracker. Participants can take a break whenever they feel fatigued. Typically, participants take a break after reading 3-5 articles. If they chose to take a break, there would be a re-calibration before the collection was restarted, considering that the person's sitting posture had a great impact on the accuracy of eye tracking. The average reading time per article is 2.88 minutes ($SD = 1.01$), which means the continuous reading time ranges from 8 to 14 minutes after one calibration.

Following the previous work~\cite{gaze-text_garain_2017}, we separated the collection of gaze data and the labeling of unknown words in order to avoid mouse clicks from affecting the user's normal eye movement behavior when reading. Participants read each article twice. During the first pass, eye movement data were collected while participants were reading. In the second pass, participants were asked to mark the unknown words they encountered in the first pass, and eye movement data were not collected for the second pass. Participants started the second pass right after the first pass to restore the feelings of the first time and help participants recall as much as possible the unknown words they encountered in the first pass. In addition, before participants read each article for the first time, the experimenter clicked a button to download all the words on the current page along with their corresponding coordinates. During the second pass, participants were required not to zoom or move the webpage to ensure that the positions of the text remained unchanged.

\subsubsection{Data Preprocessing}
To support real-time unknown word detection, we used a sliding time window to segment the data. The length of the window is 1 second, and the overlap between the two windows is zero. The data processing includes two steps. First, we located the text the user read within 1 second using the coordination of the gaze sequence. Then, we processed the raw gaze data and text data into 3 types of information which are gaze, text, and knowledge for each word in the sliding window.

\begin{figure*}[htbp]
  \centering
  \includegraphics[width = 0.9\linewidth]{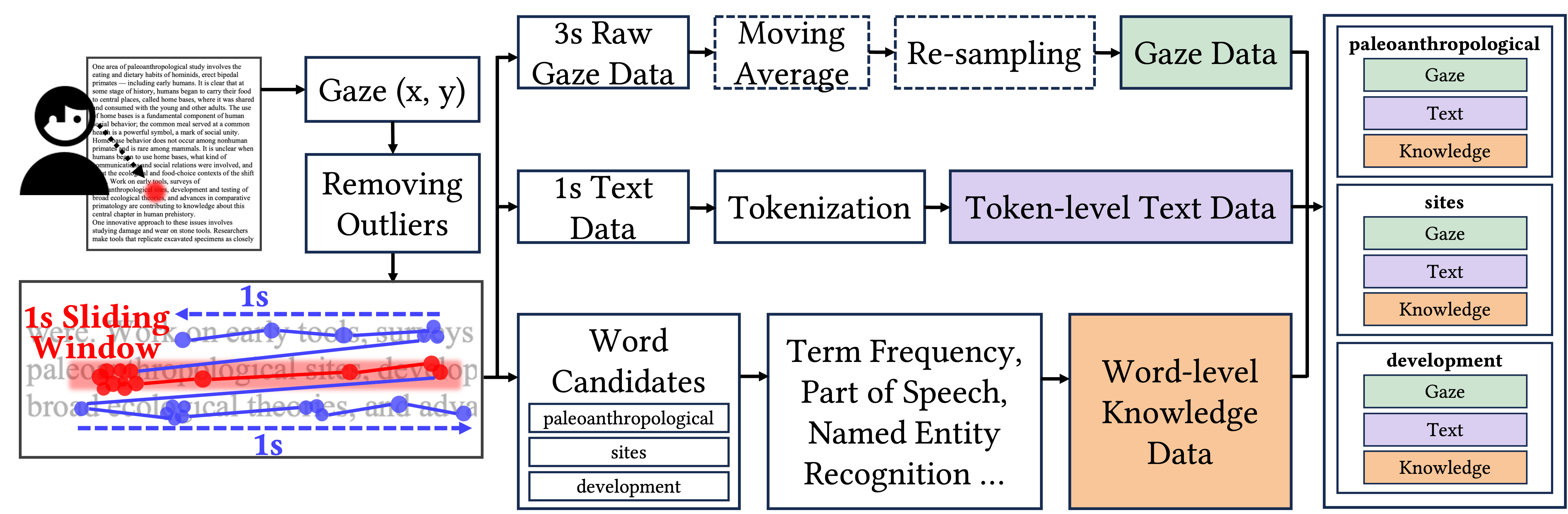}
  \caption{A bounding box is derived from the gaze coordination within a 1-second sliding window. The gaze data, token-level text data, and word-level knowledge data are calculated for each candidate word in the bounding box.}
  \label{fig:preprocessing} 
\end{figure*}

As shown in the left part of Fig.~\ref{fig:preprocessing}, we firstly de-noised the gaze data to locate the content the user is reading by getting the bounding box of the gaze coordination. Blinking can cause sudden changes in gaze data in the y direction, which will cause the bounding box to be abnormally large. We removed these outliers to avoid the extra-large bounding boxes. We analyzed the distribution of the range of the y-coordinates of the gaze data within 1 second. When the distance between a small subset of the data in a window and the other part of the data exceeds the width of three lines, we removed the smaller portion of the data. If the y-coordinates of all data in the window fluctuated greatly, we ignored this window.

After getting the bounding box of 1s gaze (red dotted box in Fig.~\ref{fig:preprocessing}), all the words except the function word (such as articles and conjunctions) in the box were regarded as the word candidate. For gaze data, the 1-second window was extended to the 3-second window by adding one second before and after the window. The consideration for this extension is that the word identification span is about 7-8 characters~\cite{RAYNER19953} and the average reading speed of participants is 2.51 words per second. We applied the moving average filter and re-sampled the gaze data collected by webcam but not eye tracker, because webcam-based data is more noisy and the sampling rate of it varies from 24Hz to 27Hz. For text data, we tokenized the content within the 1-second window. We also added prior knowledge such as the term frequencies, part of speech, and named entity recognition to each candidate word. In the example shown in Fig.~\ref{fig:preprocessing}, finally we got 3 samples from this 1-second window which are 3 words with their gaze data, token-level text data, and word-level knowledge. Labels of these words are directly derived from the mouse click file.

Compared with the data collected with eye trackers, the data collected with the webcam is noisier, and the data quality is more affected by the user's sitting posture. Therefore, we aligned the data of the first article read by the user in each session with the coordinates of the article and applied the parameters of this alignment to other articles in this session.

%% file: sections/4-experiment.tex
\section{Experiment}
\label{sec:experiment}

\change{In this section, we systematically evaluated our model. First, we present the results of our model on eye tracker data and webcam data.
Second, we analyzed the contributions of gaze and the pre-trained language model (PLM) separately.
Finally, we discuss several important aspects for practical applications, including the model's generalizability, latency, and memory consumption.}

\subsection{Experiment Settings}
\subsubsection{Training and Evaluation of \name{}}
The statistics of our collected dataset are shown in Table~\ref{tab:stat}. In the default setting (main model), the data from all users and all articles are mixed and divided into training set, development set and testing set according to 8:1:1. We fully fine-tuned our model for 30 epochs with a batch size of 32 on our dataset. The learning rate of the positional encoder-decoder model is set as 1e-3, while the rest of the model's learning rate, including the RoBERTa backbone, knowledge embeddings, and the classifier are set as 2e-5. The sample rate of the user's gaze is set as 60 Hz, with a maximum length of 3 seconds, while the maximum number of context tokens is set as 64. 

\input{tables/stats}

We report word-level accuracy, precision, recall, and F1-score in our experiments. In detail, a word is recognized as unknown if there exists at least one token within it predicted as positive. We report the F1 score on the test dataset, after selecting the best model based on the F1 score on the dev set throughout training with early stopping, with the binary classification threshold searched between 0 to 1 with the step of 0.01. The result is shown in Table~\ref{tab:main-results}.

\input{tables/main_results}

\subsubsection{Baseline}
We compared our method to an SVM baseline implemented according to~\cite{gaze-text_garain_2017} and three simple baselines. For distance and fixation heuristics baseline, the distance and fixation time calculation follow those described in Section~\ref{sec:model_positional_data_encoding}. The only difference is that we use the bounding box of words instead of tokens in this case. The logistic regression model is trained with these two features above, plus word term frequency, part of speech, and named entity tags. Results are shown in Table~\ref{tab:main-results}.

When implementing the SVM baseline, we used four gaze features (number of forward gaze points, number of backward gaze points, duration of forward reading, duration of backward reading) plus two linguistic features (word length, word frequency) mentioned in~\cite{gaze-text_garain_2017}. Because the articles we chose are longer than those in previous work, there were lots of rereads, making it difficult to determine whether it is a new line or not based on sudden changes in the x value. In addition, due to the smaller line spacing, correcting the line number based on the y value is impossible. Therefore, we only assigned line numbers to the gaze based on which line was closest to the gaze coordination and used a median filter to remove outliers. When choosing the parameters for SVM, we did the grid search and used the best one for the evaluation. The result we reported was trained using SVM (RBF kernel, C=0.01, gamma=0.1) with a class weight ratio of 6:1 between positive and negative classes.

\subsubsection{Ablation Study}
As shown in Table~\ref{tab:ablation}, we conducted ablation studies on our method by removing the contextual encoding (the use of PLMs), gaze encoding (the positional encoder-decoder), and the knowledge embeddings, respectively. When ablating the text encoding component (the PLM), we remove the text information in the inputs and discard the text encoding in our model. The model should predict the users' unknown words based on their gaze trajectories' relationship to texts' positions plus prior knowledge. When the gaze encoding is ablated, corresponding positional data and modules in Fig.~\ref{fig:model} are removed. Similarly, knowledge-oriented inputs and embeddings will be removed when we do knowledge embedding ablation. As these ablations directly remove corresponding components in our model, each would reduce our model's parameter size. To further demonstrate our model succeeds in unknown word detection due to leveraging pre-trained RoBERTa weights rather than simply scaling the model's size up, we add an extra ablation (``w/o pretrained RoBERTa'') where the RoBERTa model parameters are added to the model but randomly initialized.

\subsubsection{N-Gram Baseline}
\label{sec:experiment_n-gram}
Due to the overlap of unknown words between the training set and test set, the model might exhibit a shortcut by memorizing the unknown words in the training set. To examine the impact of this shortcut on model performance, we implemented an n-gram-based baseline to simulate the scenario where the model "memorizes" the unknown words. During prediction, if the current word combined with its preceding $n - 1$ words forms an n-gram appearing in the training set while also with the current word being labeled as unknown, the current word is classified as an unknown word at inference time. The larger the $n$, the more contextual information the n-gram baseline utilized. Since the degree of overlap in unknown words between the training and test sets varies across the main, cross-user, and cross-document settings, we applied the n-gram baseline ($n=1,2,3$) to each of these settings individually (Table~\ref{tab:n-gram}).

\input{tables/ablation}

\subsection{Model Performance}
\label{sec:main_result}


The main results of our experiment are shown in Table~\ref{tab:main-results}, the F1-score, precision and recall of our model that was trained and tested on data collected by eye tracker are 71.1\%, 63.3\% and 79.0\%. Our model significantly outperforms the heuristic baseline (F1-score 22.9\%), logistic regression (F1-score 23.4\%) and the previous work based on SVM~\cite{gaze-text_garain_2017} (F1-score 29.9\%). The performance of SVM is much worse than that presented in their paper. This could be caused by the smaller line space, which made it harder to assign the gaze to the word correctly. Besides, the document we used was more than two times longer than theirs, which led to a longer reading time (2.88 minutes per document and 8-14 minutes per reading session) and more posture changes of participants. This could worsen the tracking accuracy of eye tracker. Compared to their method, our method relies less on the gaze modality and PLMs also provide rich linguistic characteristics, thus improving the performance.

\begin{figure}
  \centering
  \includegraphics[width=1.0\columnwidth]{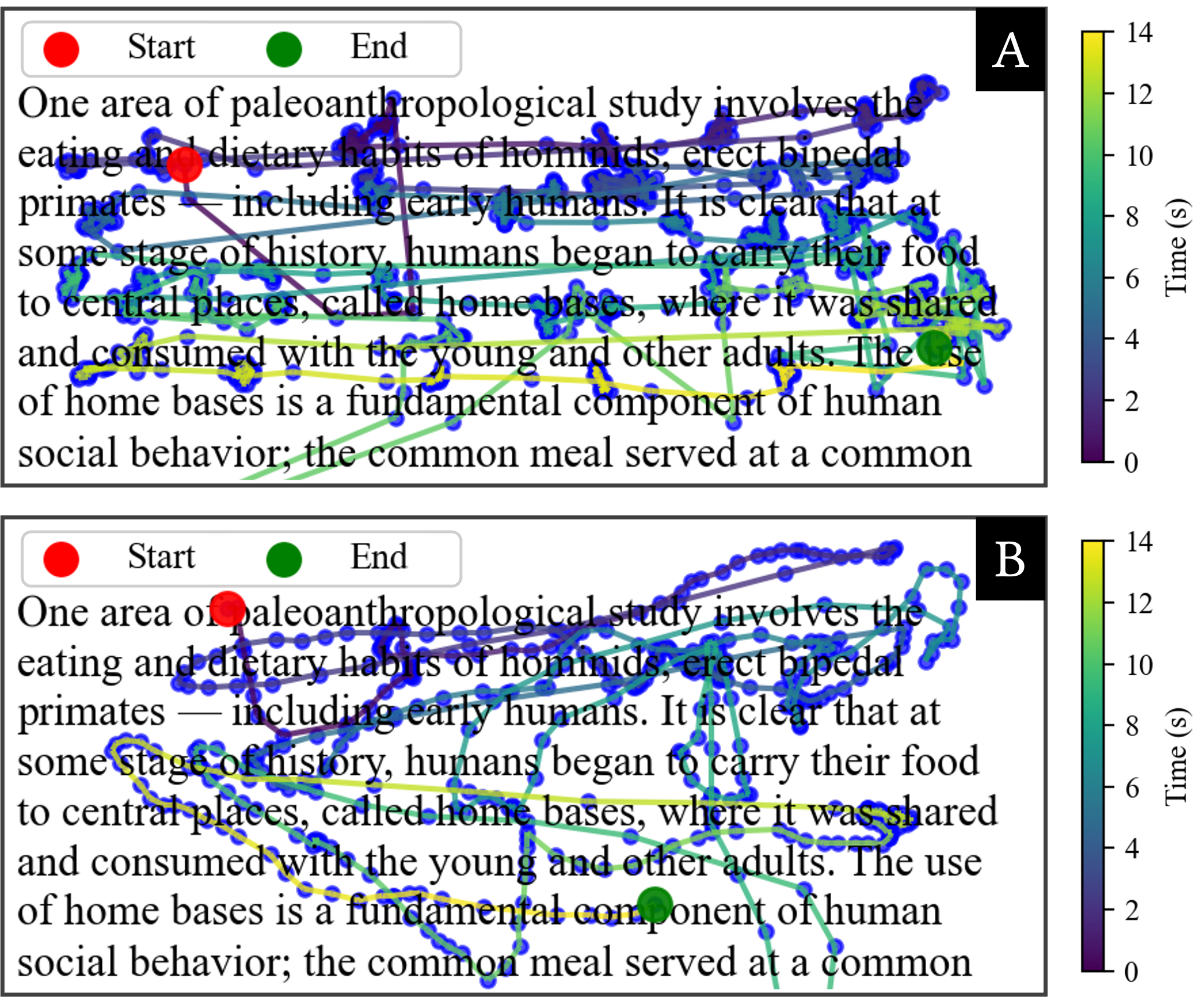}
  \caption{(A) Gaze data collected by a Tobii eye tracker. (B) Gaze data collected by a webcam.}
  \label{fig:gaze} 
\end{figure}

We also tested our model on the relatively inaccurate gaze data collected using a webcam (Fig.~\ref{fig:gaze}). To quantitatively compare the quality of gaze data collected by the eye tracker and webcam, we calculated the median absolute error (MAE) between these two types of gaze data. Since the sampling rate of the webcam is lower than that of the eye tracker, we applied cubic spline interpolation to the gaze coordination collected by webcam. The MAE in the x-direction is 115.99 pixels ($SD = 46.32$). In the y-direction, the MAE is 67.25 pixels ($SD = 32.28$). Considering that the height of the text is 20.63 pixels per line, the data from the webcam is noisier compared to the data obtained from the eye tracker. The model trained and tested on the relatively inaccurate gaze data collected using a Webcam (Fig.~\ref{fig:gaze}) also achieves a high F1-score (65.1\%) compared to baselines, even though slightly worse compared to our method trained on eye tracker collected data. This is another evidence that our method has a higher tolerance for the noise of gaze data. This result also opens the potential for more accessible solutions than previously possible.

\subsection{Contribution of Gaze}
\label{sec:contribution_gaze}

Firstly, we analyzed the differences in unknown words among users to demonstrate that the user-dependent gaze feature is necessary for correctly identifying unknown words for different users. We computed the Jaccard similarity matrix for all users. Users were divided into four groups because different groups of users read different documents. The average cross-user Jaccard similarity score is 0.24 and 77.5\% of user-paired scores are below 0.3. The low Jaccard similarity score indicates that different users have different unknown words even reading the same document. Only relying on textual information cannot identify differences between users, so it is necessary to use the user-dependent gaze features.

\begin{figure}
  \centering
  \includegraphics[width=1.0\columnwidth]{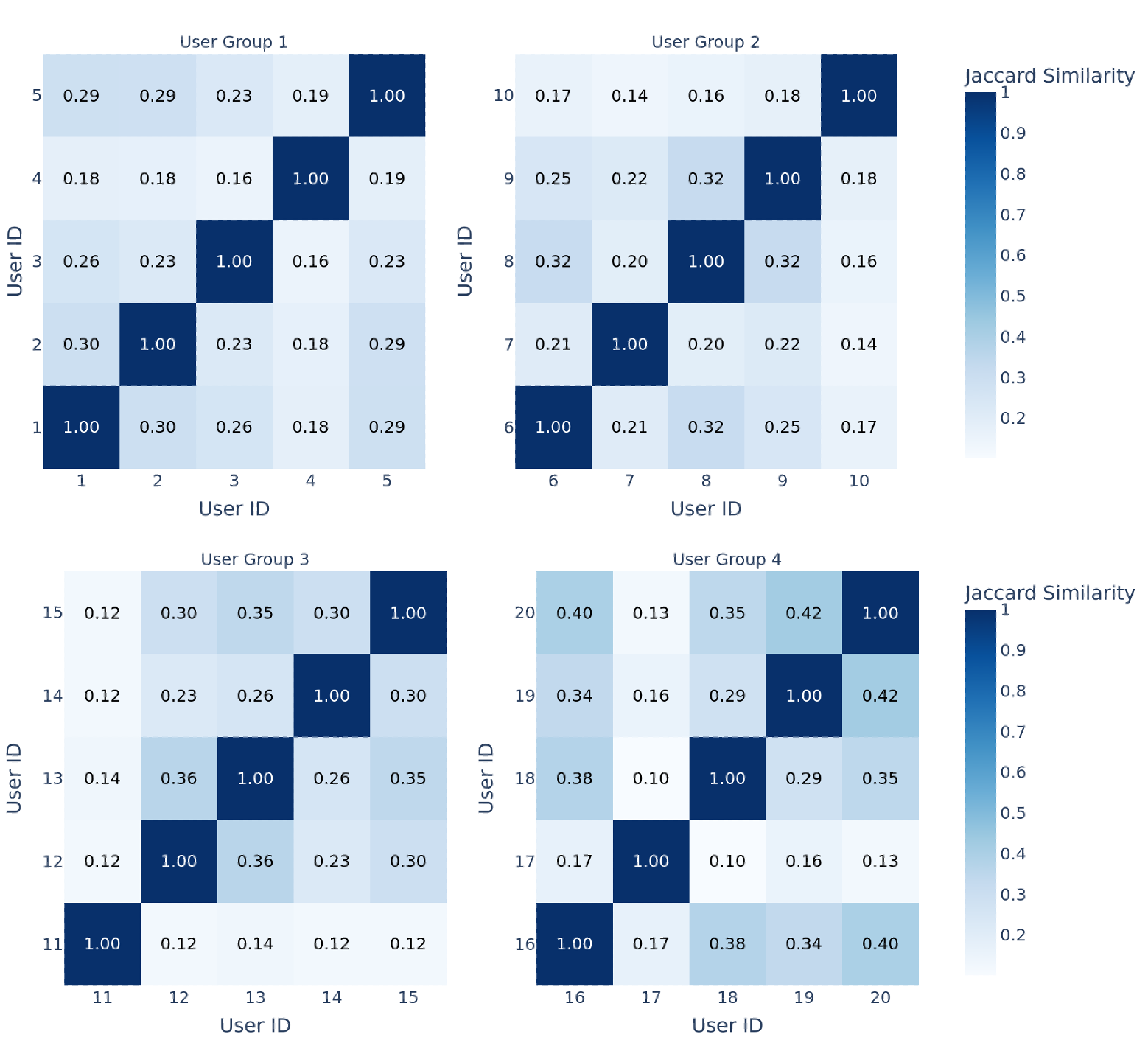}
  \caption{Jaccard similarity of unknown words between users.}
  \label{fig:similarity} 
\end{figure}

Then, we removed the gaze encoding in the ablation study to determine the extent of the contribution of gaze to unknown word detection. After removing the gaze, the recall drops from 79.0\% to 74.4\% while the precision remains almost the same. This phenomenon implies that gaze is helpful for identifying an individual's unknown words based on the user's unique behaviors and patterns.

\begin{figure}
  \centering
  \includegraphics[width=1.0\columnwidth]{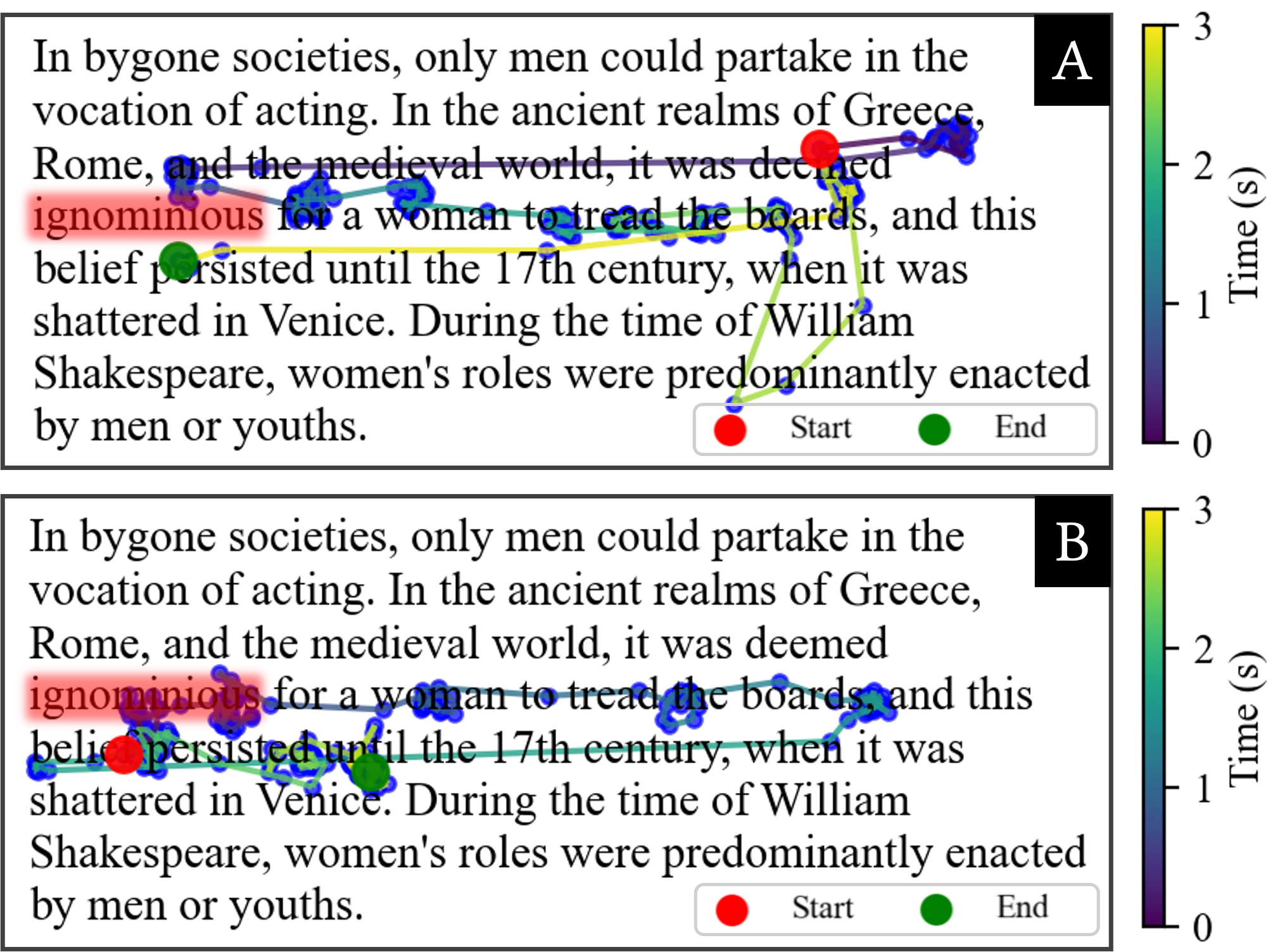}
  \caption{Case 1: (A) The word "ignominious" is not an unknown word for user A. (B) The word "ignominious" is an unknown word for user B.}
  \label{fig:case_ignominious} 
\end{figure}

We also analyzed some cases to verify that gaze can help reflect differences between users and correctly predict unknown words in real time. We plotted the clip of gaze trajectory within 3 seconds (1-second gaze to locate the region of interest and additional 2-second gaze for features) and the corresponding text in the region of interest. In Case 1 shown in Fig.~\ref{fig:case_ignominious}, the word "ignominious" is an unknown word for user B but is not an unknown word for user A. There is a noticeable dwell on "ignominious" in B's case which can facilitate the model to detect it as an unknown word correctly. The model captured this information because it predicted correctly in both cases. 

\begin{figure}
  \centering
  \includegraphics[width=1.0\columnwidth]{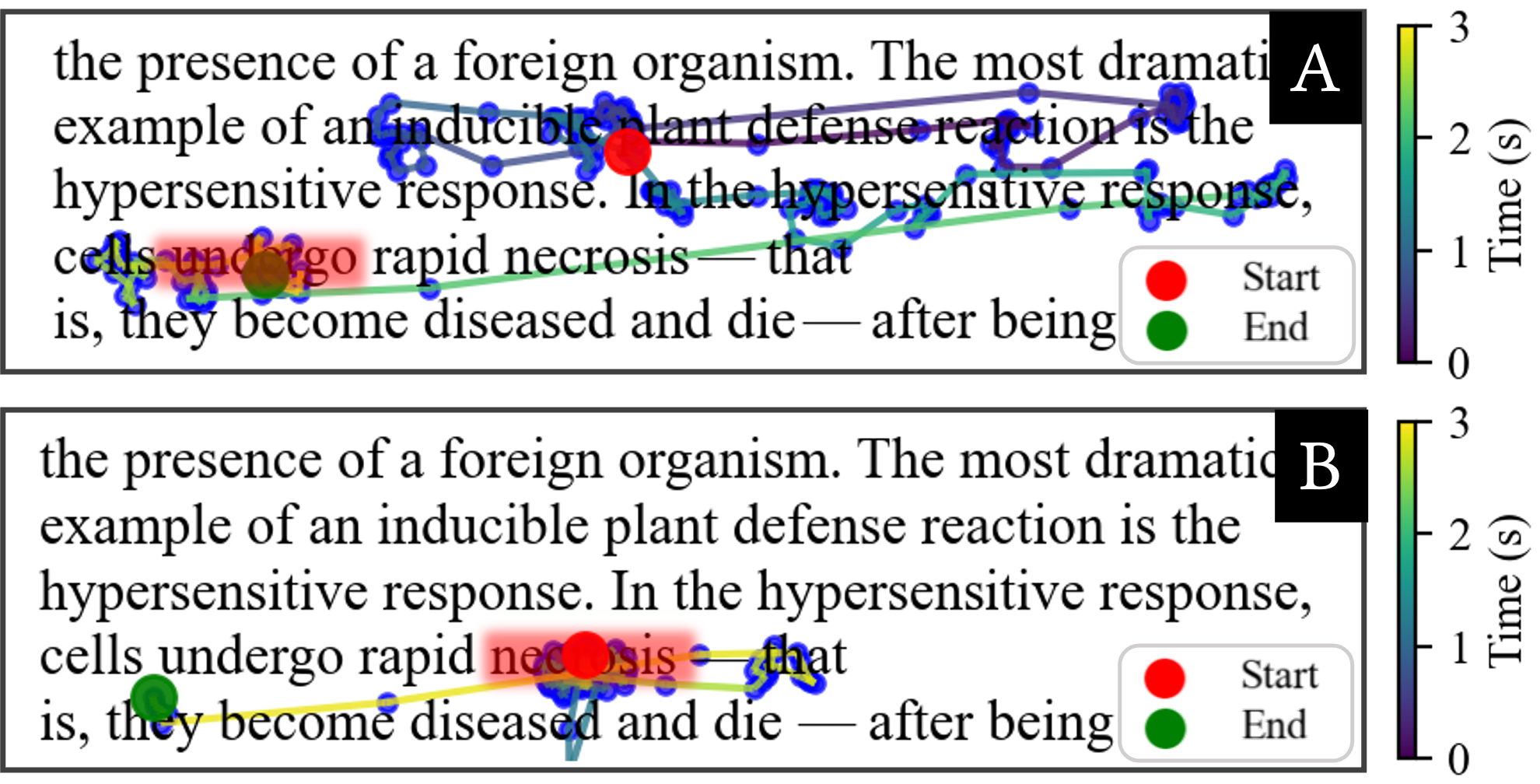}
  \caption{Case 2: (A) The word "undergo" was detected in the first second. (B) The word "necrosis" was detected in the second second.}
  \label{fig:case_2} 
\end{figure}

Another case is shown in Fig.~\ref{fig:case_2} in which the gaze in A and B are from the same user but in two consecutive seconds. The model correctly detects 'undergo' in the first second (A) and 'necrosis' in the second second (B) when the user's regions of interest in two consecutive seconds both contain 'undergo' and 'necrosis'. This result also indicates that our model learned gaze features and gaze plays an important role in real-time detection.

\subsection{Contribution of PLM}
\label{sec:contribution_PLM}

To identify the contribution of the pre-trained language model, we separately remove the "pre-trained" weights ("w/o pre-trained RoBERTa" in Fig.~\ref{tab:ablation}) by randomly initializing RoBERTa weights and remove the whole language model ("w/o textual encoding" in Fig.~\ref{tab:ablation}) by removing the RoBERTa weights. When initializing the RoBERTa randomly ("w/o pre-trained RoBERTa"), the F1-score decreases from 71.1\% to 67.6\% which indicates that the linguistic characteristics of words learned by the PLM help \name{} to detect unknown words more accurately. After entirely removing the RoBERTa parameters ("w/o textual encoding"), the F1-score significantly dropped to 22.0\%, proving the effectiveness of using PLMs for better capturing the contextual information of the documents in the task of unknown word detection. The reason why the F1-score of "w/o pre-trained RoBERTa" is relatively high compared to "w/o textual encoding" is that the number of parameters remains constant in this case though the initialization is random, allowing RoBERTa to learn linguistic features incrementally during training.

Due to the overlap of unknown words between the training and test sets, we implemented an n-gram baseline as another naive baseline to simulate the scenario where the model 'memorizes' the unknown words to analyze the effect of this shortcut. As shown in Table~\ref{tab:n-gram}, the results are best when $n=2$, with an F1 score of 51.9\%. This result is lower than our model (71.1\%), indicating that merely memorizing words is insufficient. In contrast, our model leverages the PLM to provide a more comprehensive understanding of the linguistic information from context. Moreover, another interesting phenomenon is that the F1 score of the 2-gram baseline (51.9\%) is higher than the gaze-based SVM baseline (29.9\%). This suggests that textual information contributes more than gaze data when the line space is so small that the gaze is not accurate enough to locate individual words.

\input{tables/n-gram_baseline}

We also conducted the ablation study to the cross-document setting when the overlap of unknown words between the training and test sets is very small (Jaccard Similarity is 0.006). As shown in Table~\ref{tab:cross_doc_ablation}, removing gaze leads to a drop in performance, but it is less significant compared to the drop caused by the absence of pre-training.

Synthesizing the insights from these three experiments, we conclude that gaze is indispensable for locating regions of interest and providing real-time detection, and it also improves the model performance. However, the linguistic information introduced by the PLM plays the primary role in enhancing the model performance.

\subsection{Cross-User and Cross-Document Generalizability}
\label{sec:cross-user}
We evaluate our method with the cross-user generalizability with the leave-one-user-out setting, where the data collected from 18 randomly-selected users are set as the training dataset, while the rest of the two user's data are used as the dev and test sets. We find that in cross-user settings, the F1 score of our method drops to 59.6\%, with the accuracy, precision, and recall decreasing to 97.3\%, 52.8\%, and 68.4\%. Compared to the main setting, the performance decline of the model in the cross-user setting could be attributed to the loss of personalized information due to the absence of new user's gaze data. Another possible reason is the lack of labels for unknown words from new users, which affects the model's ability to assess word difficulty accurately for them. This suggests that it is necessary to collect some data from new users for fine-tuning before deployment to better adapt the model to different users. Additionally, relying heavily on the PLM might hinder the model's ability to achieve personalization.

\input{tables/cross_user_doc}
To demonstrate that \name{} can indeed detect users' unknown words instead of simply memorizing difficult words in the documents, we also evaluate our model with the leave-one-document-out setting. In this case, we equally split the 120 reading documents into groups of 10, and in each training run, we trained our model on 110 documents while testing the model on the remaining 10 documents. The average Jaccard similarity score between the training set and test set is 0.006, demonstrating that the unknown words in the test set are very different from those in the training set. Our model's F1 score drops to 51.1\%, with the precision, recall, and accuracy shifts to 43.3\%, 62.8\%, and 96.8\%, respectively. The result indicates that generalizing the capabilities of unknown word detection across documents is challenging because of the lack of text context. Since the difference in context between the training and test set in the cross-document setting is larger than that in the cross-user setting, the performance in the cross-document setting is worse than cross-user setting. However, our model still achieves non-trivial improvements compared to the random classifier (F1-score 11.1\%) and n-gram baseline (F1-score 7.9\%), indicating that \name{} learns essential user gaze patterns and linguistic knowledge through training beyond naively memorizing difficult words.

To further investigate how gaze and PLM impact model performance, we conducted an ablation study in this challenging cross-document setting. After removing gaze and pre-training, the F1-score decreased to 49.9\% and 37.4\%, respectively. This result indicates that gaze contributes to the model's performance, but its impact is smaller compared to the contribution of textual information. Combined with the fact that the n-gram baseline's F1-score is only 7.9\%, it can be concluded that the contribution of textual information lies not in merely memorizing words but in enhancing the model's understanding of context.

\input{tables/cross_doc_ablation}

\subsection{Latency and Memory Consumption}
To demonstrate the capabilities of our method being used in real-time applications, we evaluate our method's latency with both CPU and GPU usage during inference. We set the batch size as 1 during inference latency testing. In GPU usage testing, we test the model with an RTX 4090 graphic card, where the average inference latency of the model is 0.013 seconds. Meanwhile, in CPU usage testing, the model's latency is 0.036 seconds. Overall, these results indicate that our method can support real-time applications with latency within 1 second. Moreover, under both settings, the maximum memory consumption of our model is 488.41MB, further proving that our method can be easily adapted to different on-device settings, widening our method's practicality for various downstream application supports.

%% file: tables/stats.tex
\begin{table*}[]
\caption{Statistics of our collected data.}
\label{tab:stat}
\begin{tabular}{@{}rrrrrrl@{}}
\toprule
\#Users & \#Documents & \#Train Data & \#Dev Data & \#Test Data & \#Tokens & \#Unknown Tokens           \\ \midrule
20      & 120         & \change{9,802}         & 980        & 980         & 380,524  & \multicolumn{1}{r}{25,233} \\ \bottomrule
\end{tabular}
\end{table*}

%% file: tables/main_results.tex
\begin{table*}[]
\caption{The Main Results of our method compared to heuristic and logistic regression baselines, backed with eye tracker and webcam collected user gaze data. Our model on the eye tracker data performs the best (highlighted in bold).}
\label{tab:main-results}
\begin{tabular}{@{}llrrrr@{}}
\toprule
Device      & Method              & Accuracy (\%) & F1 (\%)   & Precision (\%) & Recall (\%) \\ \midrule
Eye tracker & Distance heuristics & 76.6     & 20.6 & 13.3      & 45.5   \\
Eye tracker & Fixation heuristics & 82.4     & 22.9 & 16.2      & 39.1   \\
Eye tracker & Logistic regression & 96.6     & 23.4 & 15.9      & 44.5   \\
Eye tracker & SVM~\cite{gaze-text_garain_2017} & 83.0 & 29.9 & 20.3 & 56.7 \\ \midrule
Eye tracker & Ours (main model)                & \textbf{97.6}     & \textbf{71.1} & \textbf{63.3}      & \textbf{79.0}   \\
Webcam      & Ours (main model)                & 97.3     & 65.1 & 60.3      & 69.7   \\ \bottomrule
\end{tabular}
\end{table*}

%% file: tables/ablation.tex
\begin{table*}[]
\caption{Ablation study with eye tracker collected gaze data.}
\label{tab:ablation}
\begin{tabular}{@{}rrrrr@{}}
\toprule
\multicolumn{1}{l}{Method}     & Accuracy (\%) & F1 (\%)   & Precision (\%) & Recall (\%) \\ \midrule
\multicolumn{1}{l}{Ours (main model)} & 97.6     & 71.1 & 63.3      & 79.0   \\ \midrule
w/o textual encoding        & 96.6     & 16.2 & 8.1      & 91.4   \\
w/o gaze encoding              & 97.5     & 68.5 & 63.5     & 74.4  \\
w/o knowledge embedding        & 97.6     & 69.2 & 68.4    &  70.0  \\
w/o pretrained RoBERTa        & 97.7     & 67.6 & 61.9    &  74.6  \\ \bottomrule
\end{tabular}
\end{table*}

%% file: tables/n-gram_baseline.tex
\begin{table}[]
\caption{N-gram baseline (n=1, 2, 3) for main, cross-user and cross-document settings.}
\label{tab:n-gram}
\begin{tabular}{lrrr}
\toprule
Method            & F1 (\%) & Precision (\%) & Recall (\%) \\ \midrule
Main 1-gram       & 23.8    & 14.0           & 79.9        \\
Main 2-gram       & 51.9    & 38.5           & 79.4        \\
Main 3-gram       & 30.8    & 29.0           & 32.9        \\ \hline
Cross-user 1-gram & 19.3    & 11.4           & 62.5        \\
Cross-user 2-gram & 39.9    & 31.3           & 55.7        \\
Cross-user 3-gram & 19.1    & 16.9           & 22.1        \\ \hline
Cross-doc 1-gram  & 7.9     & 4.9            & 21.0        \\
Cross-doc 2-gram  & 5.9     & 9.0            & 4.5         \\
Cross-doc 3-gram  & 3.5     & 4.8            & 2.8         \\
\bottomrule
\end{tabular}
\end{table}

%% file: tables/cross_user_doc.tex
\begin{table}[]
\caption{Cross-user and Cross-document generalizability.}
\label{tab:cross}
\begin{tabular}{lllll}
\toprule
Method     & Accuracy (\%) & F1 (\%) & Precision (\%) & Recall (\%) \\ \midrule
Main       & 97.6          & 71.1    & 63.3           & 79.0        \\
Cross-user & 97.3          & 59.6    & 52.8           & 68.4        \\
Cross-doc  & 96.8          & 51.1    & 43.3           & 62.8        \\ \bottomrule
\end{tabular}
\end{table}

%% file: tables/cross_doc_ablation.tex
\begin{table*}[]
\caption{Ablation study on leave-one-document-out setting.}
\label{tab:cross_doc_ablation}
\begin{tabular}{@{}rrrrr@{}}
\toprule
\multicolumn{1}{l}{Method}     & Accuracy (\%) & F1 (\%)   & Precision (\%) & Recall (\%) \\ \midrule
\multicolumn{1}{l}{Cross-document} & 96.8     & 51.1 & 43.3      & 62.8   \\ \midrule
w/o textual encoding        & 96.7        & 18.9    & 11.4        & 65.0   \\
w/o gaze encoding              & 96.8     & 49.9 & 42.8     & 60.1  \\
w/o knowledge embedding        & 96.7     & 34.0 & 26.6    &  49.6  \\
w/o pretrained RoBERTa        & 96.7     & 34.7 & 26.6    &  51.0  \\ \bottomrule
\end{tabular}
\end{table*}

%% file: sections/5-evaluation.tex
\section{User Evaluation}
\label{sec:user_evaluation}
The user experience using the application built on \name{} depends on the real-time performance of \name{}. To evaluate the performance of our method in a real-time scenario, we built a reading assistance prototype that provided the translation and explanation of words while the user was reading. Then, we calculated the metrics such as F1-score and reading time of \name{} and also analyzed the subject feedback from users.

\subsection{User Study Design}
\label{sec:user_study_design}
The questions we want to know the answers are
\begin{enumerate}
    \item What is the real-time performance (F1-score, correctly triggered rate, and false alarm rate) of our unknown word detection method (Section~\ref{sec:eval_realtime_performance})?
    \item Can our method make reading more fluent (Section~\ref{sec:eval_reading_fluency})?
    \item How is user experience when comparing our system to the commonly used method (Section~\ref{sec:eval_user_experience})?
\end{enumerate}

To answer these questions, we implemented three methods for real-time unknown word detection and compared them regarding objective metrics and subject scale. The first method (\name{}) uses gaze to locate the region of interest and applies our model to detect unknown words. When a new word is detected, its translation and explanation will be automatically displayed in the sidebar, and users do not need to take any action. The second method (Click) imitates the typical process of looking up words by clicking. When a user selects or double-clicks a word, a "meaning" option pops up for translation and explanation of the word.  The third method (Ideal) simulates an ideal scenario where the model is highly accurate. Before users start reading, they are required to select unknown words from a list of candidate words. As users start reading, the words in the region of interest are compared with the selected unknown words. The system then displays the words in the unknown word list. To control variables other than prediction accuracy, all data processing and model inference procedures for Ideal are the same as those of \name{}. This approach aims to eliminate any degradation in user experience caused by model inaccuracies and to solely evaluate the potential of the unknown word detection method using gaze and text.

For the objective metrics, we calculated the accuracy, F1-score, correctly triggered rate, and false alarm rate for \name{}. We also recorded the reading time of users under three methods.

For user experience, we conducted the evaluation for the three reading assistance based on each method across the following five aspects using a 5-point Likert scale:
\begin{itemize}
    \item Preference: To which degree do you prefer the application? 1 for ``I prefer this one least''. 5 for ``I prefer this one most''.
    \item Willingness to use: Would you like to use it in the future? 1 for ``I won't use this application in the future''. 5 for ``I would definitely use the application if it were available.'' 
    \item Usefulness: How helpful do you think it is for your reading and vocabulary learning? 1 for ``Not helpful at all''. 5 for ``Very helpful''.
    \item Perceived reading fluency: How do you feel about your reading fluency when using this feature? 1 for ``My reading is still slow and unsmooth''. 5 for ``It speeds up my reading a lot and makes my reading very fluent''.
    \item Perceived latency: What do you think of the latency from when you need help to when the system pops up an explanation? 1 for ``The latency is very small and does not affect usage''. 5 for ``The latency is very large and makes the system not usable at all''.
\end{itemize}
Then we compared the rate for three methods using Wilcoxon signed-rank test at .05 significance level.
    
\subsection{Setup and Procedure}
\label{sec:eval_setup}
\subsubsection{Implementation of a Reading Assistance Proof-of-Concept}
\begin{figure}
  \centering
  \includegraphics[width=1.0\columnwidth]{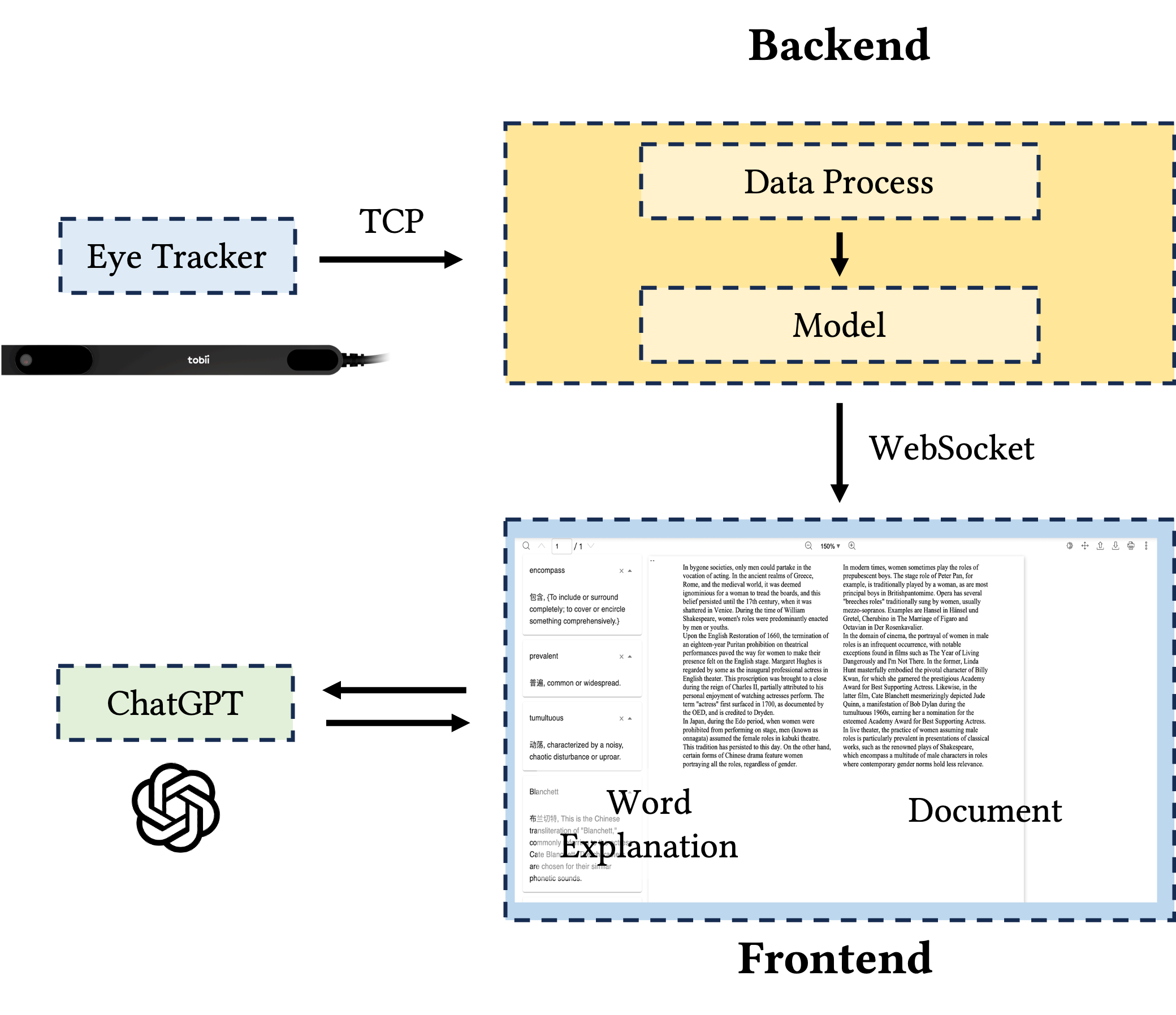}
  \caption{The implementation of reading assistance prototype based on our unknown word detection method.}
  \label{fig:PoC} 
\end{figure}

The proof-of-concept is a web-based application as shown in Fig.~\ref{fig:PoC}. The back-end was built using Python and the front-end was built using TypeScript and React. The React-PDF-viewer\footnote{https://react-pdf-viewer.dev/} was used to show the pdf. For the \name{} and Ideal, the gaze was captured by Tobii Nano which was the same model as the one used in data collection for model training. Next, the data was transmitted to the Python back-end via TCP and input into the unknown word detection model for inference along with text data. The model we used here is the main model trained on all the data and without any ablation. The same data processing step as the training was applied to gaze and text data. Then, unknown words predicted by the model were sent to the front-end through a web socket. The OpenAI API was called in the front-end to get translations and contextual explanations of words. In the end, the translations and explanations of words were displayed on the web page's sidebar. For the Click, the unknown words were detected by a click listener and then called the OpenAI API to get the translations and explanations. The study was conducted on a MacBook Pro (CPU: Apple M1 Pro, RAM: 16G, screen: 14 inches).

\subsubsection{Participant and Material}
\label{sec:eval_participant}
We recruited 10 English as a second language learners aged between 22-28 ($M=23.80, SD=1.87$), including 4 males and 6 females. Participants in this evaluation are different from participants for data collection. Eight of them wore glasses and two of them didn't wear glasses. Their Vocabulary Levels Test (VLT) score on the 5000-word frequency level group ranged from 10 to 29 ($M=18.80, SD=5.07$).

The documents we used are from TOEFL reading material with similar lengths (390 words/24 lines, 353 words/23 lines, 371 words/22 lines). Another important reason for selecting these three documents is that, their total numbers of unknown words (49 words, 49 words, 52 words) marked in the previous data collection are close. During the user study, the text was displayed in two columns, consistent with the settings used for data collection. We employed Times New Roman font at size 10 with single line spacing.

\subsubsection{Procedure}
Firstly, the experimenter introduced the background of the study and randomly assigned three articles to three tasks (\name{}, Click, and Ideal). Afterward, the eye tracker was calibrated and the participants were required to try the \name{} and Click to get familiar with them. After the calibration, participants were informed to maintain the same posture as much as possible and read with their original reading habits. 

Participants used \name{} when reading the first document. The application automatically saved the unknown words detected by the model, the user's reading time, and gaze data. Participants marked unknown words they encountered for ground truth after the reading. Participants used Click and got word explanation by clicking the word when reading the second document. The reading time was recorded. For the Ideal, participants first deleted words they knew from the unknown word candidate list before the reading. The gaze function was turned to locate the region of interest and the explanations for words from the unknown word list were displayed when the participant encountered those words. Reading time and gaze data were recorded. After reading three documents, the participants were asked to rate \name{}, Click, and Ideal individually in the first aspects mentioned in Section.~\ref{sec:user_study_design}.

\subsection{Analysis and Results}
\label{sec:eval_result}
\subsubsection{Real-Time Performance}
\label{sec:eval_realtime_performance}
We calculated the F1-score, precision, and recall for each participant by comparing the unknown words marked by the participants and detected by the model in \name{} setting. The averaged F1-score, precision, and recall are respectively 56.54\%, 51.93\%, and 67.39\%, which is close to the cross-user result in Section~\ref{sec:cross-user} (59.6\%, 52.8\%, and 68.4\%). This result shows that our model performs the same when running offline and in real-time, which proves that our model can work in real-time. The average correctly triggered rate is equal to recall which is 67.39\% and the average false alert is 2.89\%. The false alert rate is similar to the previous work conducted using the more accurate head-mounted eye tracker~\cite{idict2006hyrskykari}. The low correctly triggered rate could be caused by the inaccuracy of our model and the inevitable change of posture when the user was reading.

\subsubsection{Reading Fluency}
\label{sec:eval_reading_fluency}
We evaluated the influence of our word detection method on improving users' reading fluency from two perspectives which are reading time and user perceived fluency. The reading time of \name{} ($p = 0.006$) and Ideal ($p = 0.048$) is significantly shorter than the reading time of Click. There is no significant difference between the reading time of \name{} and Ideal ($p = 0.193$).

For the perceived reading fluency, the rate of ideal is significantly higher than Click ($p = 0.030$), but there is no significant difference between \name{} and Click ($p = 0.0773$). Automatically detecting unknown words speeds up the reading by minimizing the number of operations. However, participants' feedback shows that the explanation of incorrect unknown words that popped up in \name{} distracted users and slowed down the reading when using \name{}. Ideal does not have this problem and will not interrupt reading due to mouse operation, so it makes the reading process more fluent.

\begin{figure}[ht]
    \centering
    \includegraphics[width=0.49\columnwidth]{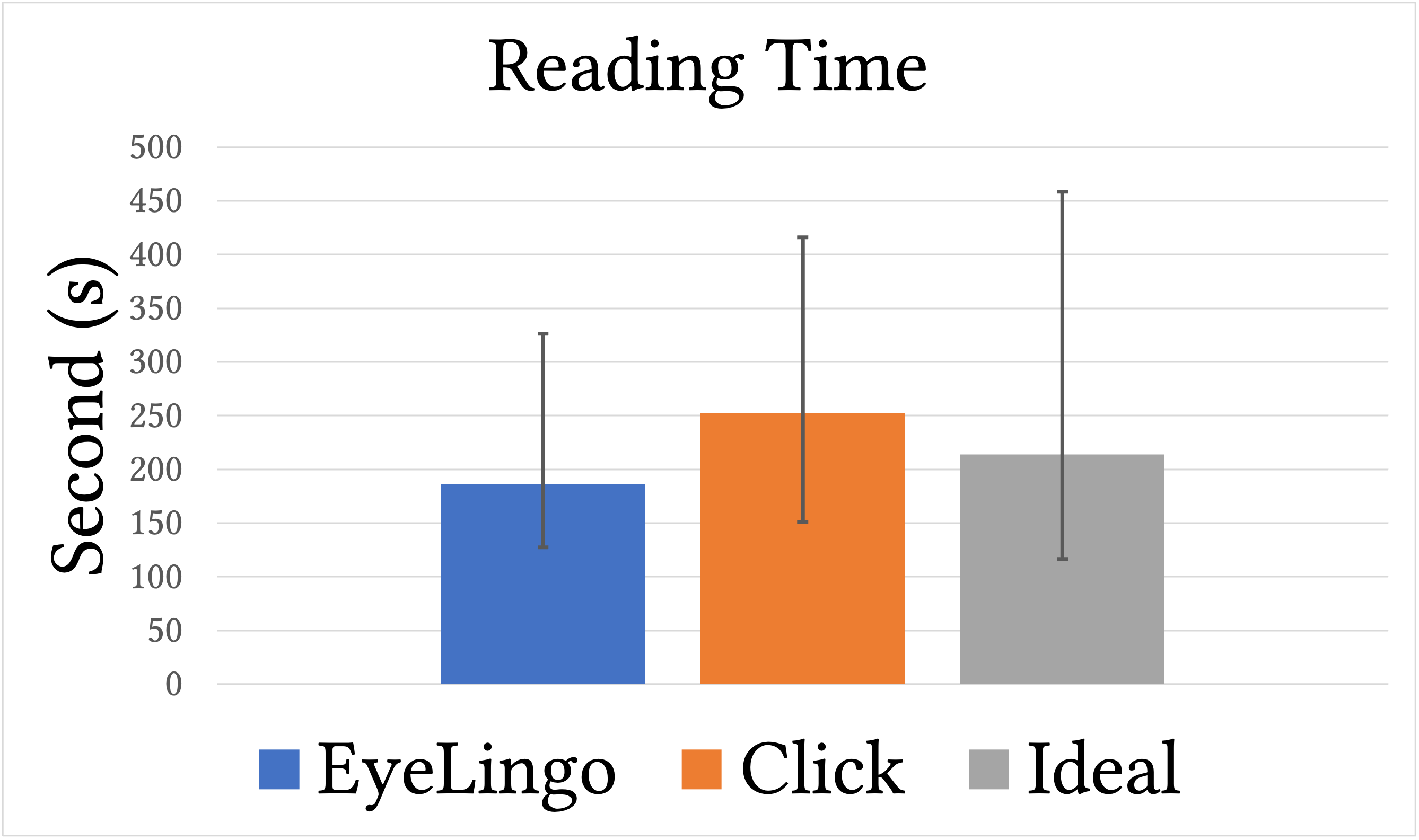}
    \includegraphics[width=0.49\columnwidth]{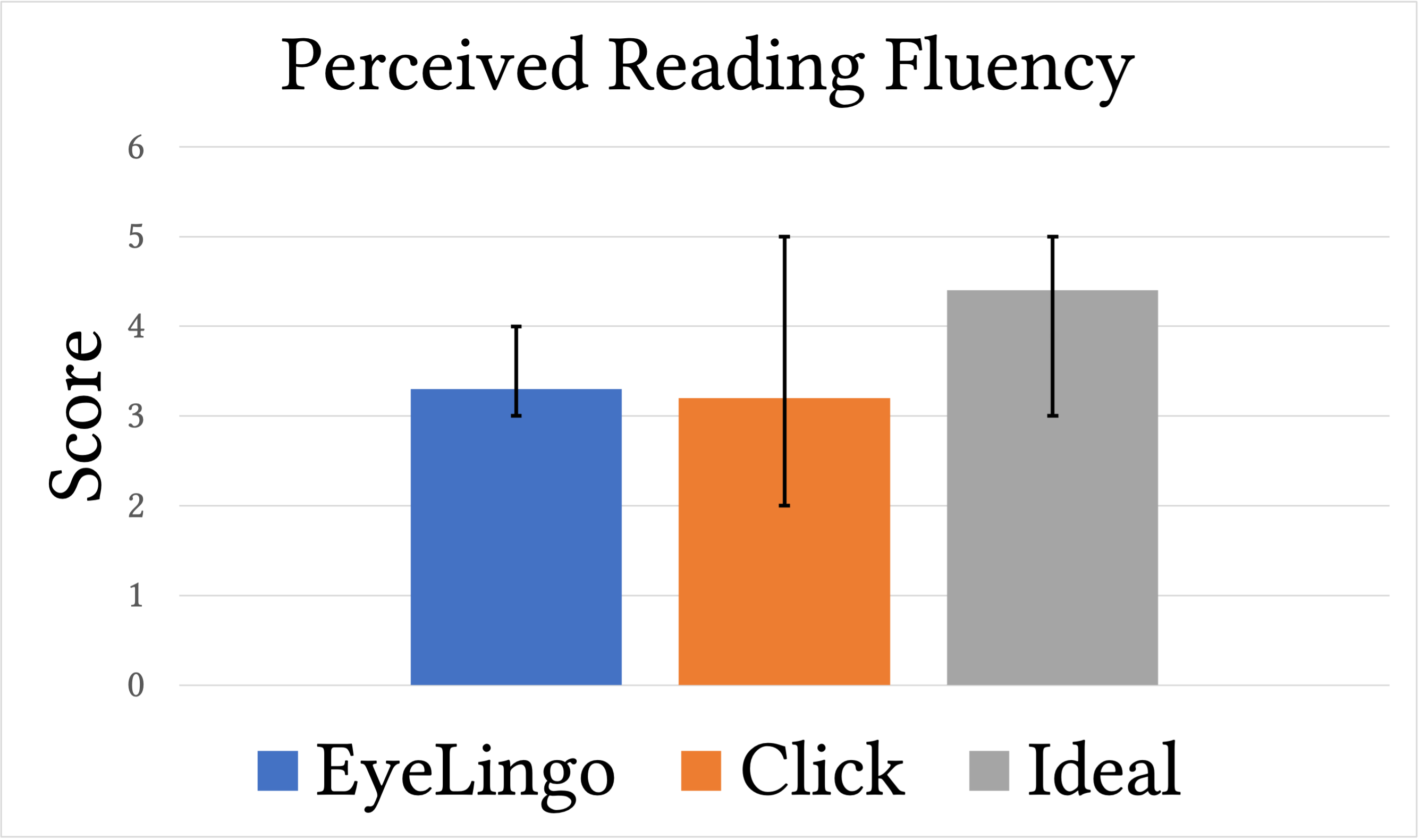}
    \caption{\name{} enhances the fluency of reading for users compared with getting word explanation by clicking the unknown word.}
    \label{fig:eval_fluency}
\end{figure}

\subsubsection{User Experience}
\label{sec:eval_user_experience}
\begin{figure}[ht]
    \centering
    \includegraphics[width=0.49\columnwidth]{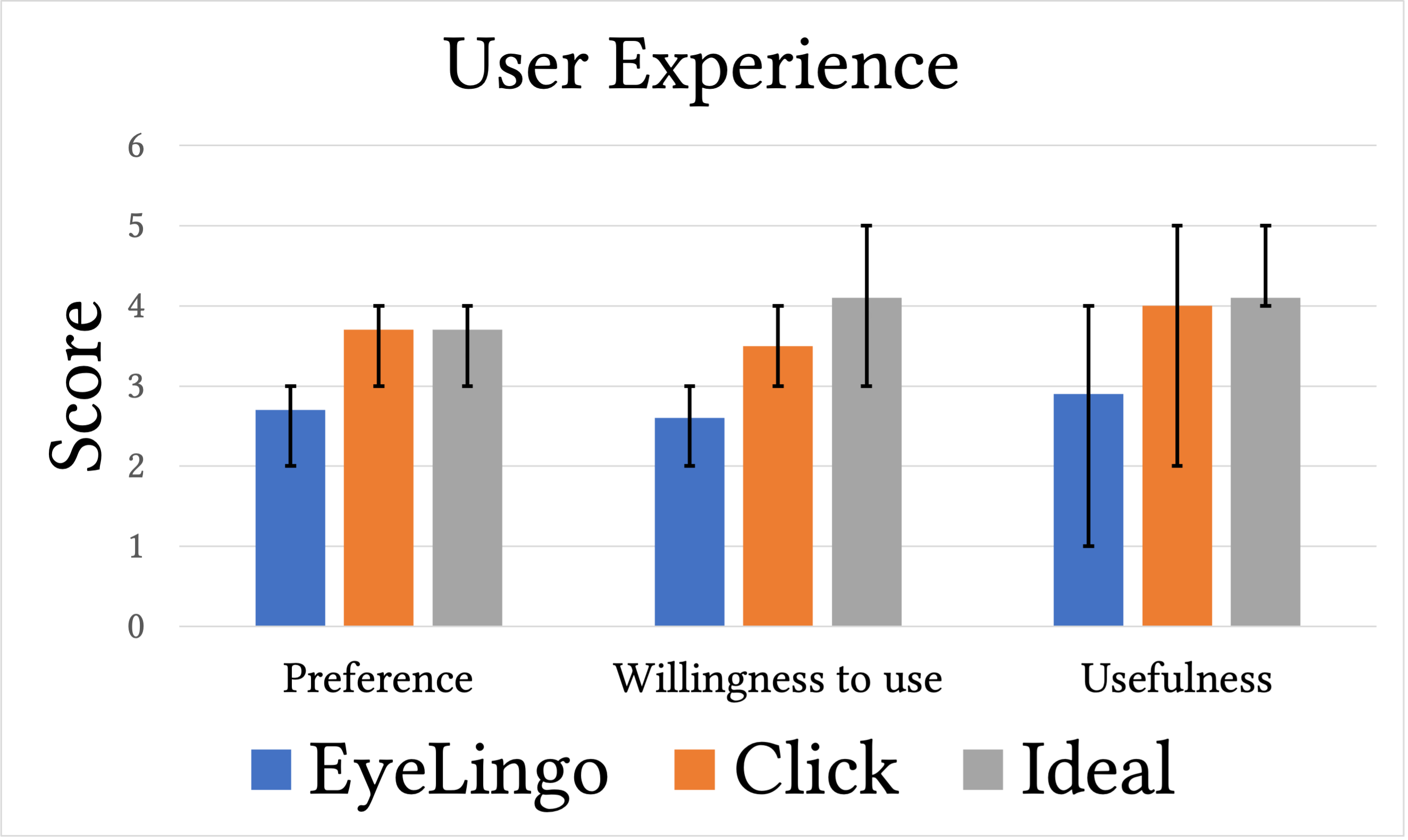}
    \includegraphics[width=0.49\columnwidth]{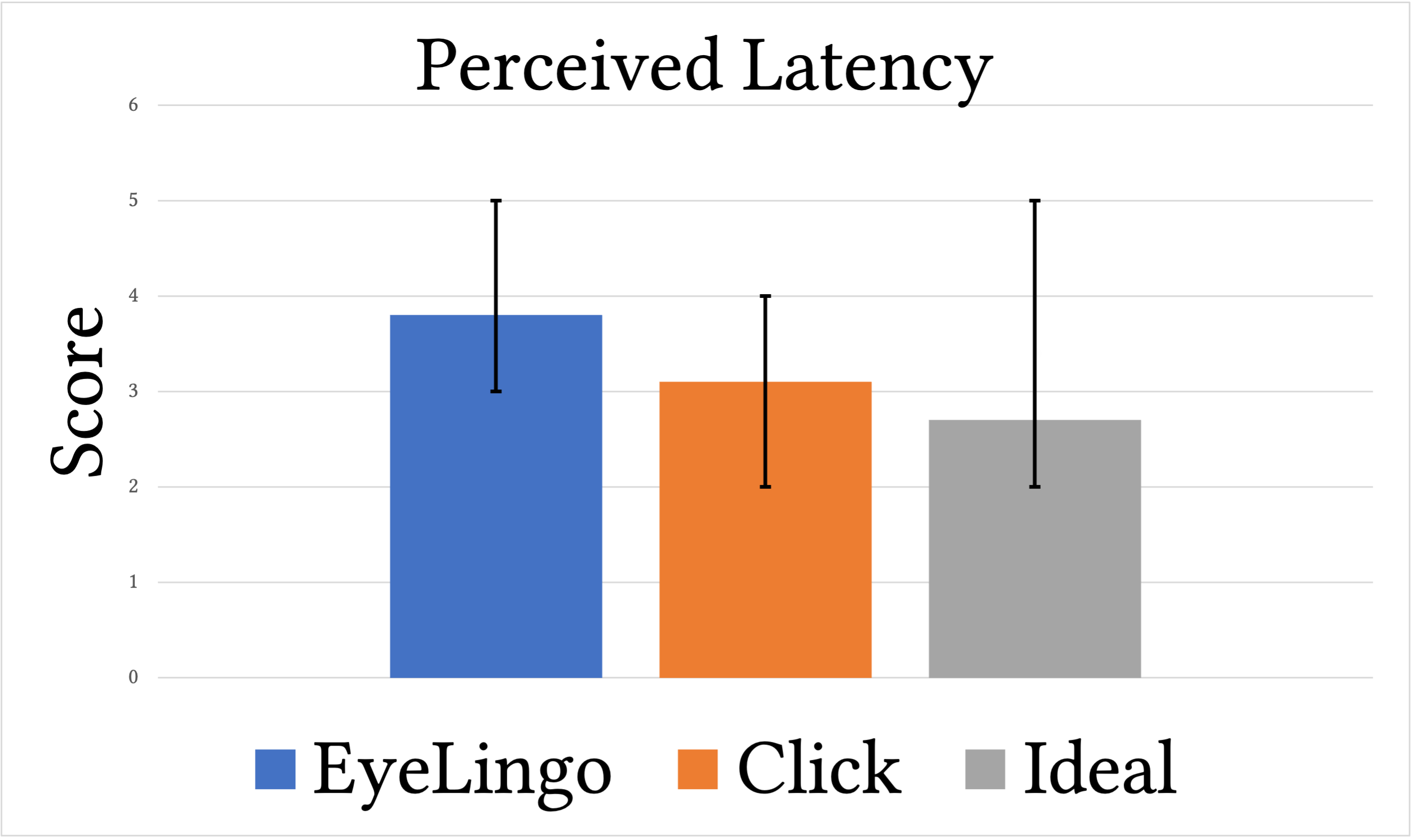}
    \caption{Subjective rating of user experience. Left: the ideal condition of our method improves the willingness to use and usefulness compared to the click method. Right: the latency that participants perceived of ideal condition is less than click.}
    \label{fig:eval_subjective}
\end{figure}

For user preference, the average scores for Click and Ideal are the same as shown in Fig.~\ref{fig:eval_subjective}. It indicates that detecting unknown words by gaze and PLMs has the potential to replace the conventional method of word lookup via mouse clicks. In terms of willingness to use, the score of Ideal (4.1) is higher than Click (3.5). Most participants stated they preferred real-time automatic word detection if the accuracy was high. For the usefulness, the score of Ideal (4.2) is slightly higher than Click (4.0). Participants indicated that our proposed method helps them read more fluently with less disruption. The perceived latency of Ideal (2.7) is lower than Click (3.1) demonstrating that the latency caused by the data processing and model inference is negligible because the Ideal also includes these components from \name{}. Moreover, the time saved by avoiding clicks can compensate for the delay introduced by GPT, thereby reducing overall latency.

The scores of \name{} are lower than Click, but Ideal can surpass traditional click methods in most aspects. Considering that the major difference between \name{} and Ideal is the accuracy of the unknown word detection model, \name{} could offer a more natural and efficient interaction mechanism by further improving the accuracy in the future.

%% file: sections/7-discussion.tex
\section{Discussion}
\label{sec:discussion}

In this section, we first summarize the conclusion and share some key observations. Then, we reflect on the usability of our method and propose potential applications. In the end, we discuss the limitations and future work.

\subsection{Effectiveness of \name{}}
\label{sec:discuss_effectiveness}
Firstly, based on the results from Section~\ref{sec:experiment}, we can draw the following conclusions:
\begin{itemize}
    \item It is efficient to detect unknown words by combining linguistic characteristics provided by the pre-trained language model (PLM) and gaze trajectory.
    \item The prediction is mainly based on the linguistic features from the textual context captured by PLM.
    \item Gaze locates the region of interest in a timely manner, which is necessary for real-time applications. Gaze also helps improve the model performance, but its contribution is limited compared to PLM.
\end{itemize}

Additionally, it is interesting that while we typically assume that the gaze modality should contribute significantly to the task of unknown word detection, the experimental results show that the contribution of gaze to the model’s improvement is small with the existence of PLM. Based on the previous analysis of line spacing and eye tracker accuracy, a possible reason for this is that under normal reading conditions (single-line spacing, line height 3-5 mm), the eye tracker’s accuracy is insufficient to precisely detect which line the gaze belongs to, thus failing to accurately locate the gaze on the words. Furthermore, changes in user posture during long reading sessions further reduce the accuracy of the eye tracker. In our system, PLM compensates for this issue by providing linguistic information based on the text.

From another perspective, the low contribution of gaze is not necessarily a disadvantage. Our method’s reduced reliance on gaze makes it more tolerant of noise. The model’s good performance on data collected by webcams further supports this conclusion. The reduced dependency on gaze data allows our model to be applied on more affordable and accessible devices, such as webcams.

\subsection{Usability of \name{}}
\label{sec:discuss_usability}
The results from the user evaluation (Section~\ref{sec:user_evaluation}) show that our reading assistance prototype helps users read more fluently and they are more willing to use it compared to traditional click-to-translate methods. In addition to providing real-time translation and explanations during reading, our system can also benefit ESL for long-term learning. For example, based on the unknown word detected by our system, we can generate a vocabulary list for memorizing and offer memory curve tracking. Furthermore, these unknown words can also be used to generate personalized summaries and notes.

The potential issue of generalizability across users, texts and devices can be addressed through fine-tuning and reinforcement learning methods. During the initial phases of usage, the system collects both gaze and text data for fine-tuning and lets users provide feedback on the model's predictions. This allows the model to continuously learn the user's unique gaze patterns and infer their vocabulary proficiency and domain expertise from textual content, thereby improving prediction accuracy.

\subsection{Limitation and Future Works}
\label{sec:discuss_limitation}
The quality of gaze data hinders the improvement model performance. The accuracy of the eye tracker is not enough for word-level detection. Common formatting, such as single-line spacing and 10-point font, results in a line height of approximately 3-5 mm when viewed using the PDF viewer with a sidebar on a 14-inch laptop. This requires an accuracy of about $0.3-0.6^\circ$ at a reading distance of 50-60 cm. However, most eye trackers have a gaze accuracy ranging from $0.2-1.1^\circ$~\cite{gaze_survey_2024}. Combined with additional errors caused by head and upper body movements, this level of accuracy is insufficient for real-world reading scenarios. During data collection and evaluation, some participants reported that even after calibration, the error could span 1-3 lines. This makes it difficult to determine the specific word the user is focusing on based solely on gaze coordinates, explaining why gaze-based baselines performed poorly on our data.

\change{The inaccuracy of the gaze data could also lead to the inaccuracy of data labeling. To mitigate the impact of mouse clicks on gaze behavior, we asked users to label unknown words during their second pass. However, this widely adopted labeling method inherently requires "guessing" which words correspond to a given gaze trajectory. Previous works mapped each gaze coordinate directly to a specific word to establish word-gaze pairs. This method is infeasible for text with normal line spacing, so we establish gaze-word pairs by defining a bounding box based on a segment of gaze to identify the corresponding words instead. While this approach improves robustness, it may also introduce mismatches between gaze and words and thus introduce noise to the dataset. To further improve model performance, more precise labeling methods are needed.}

Additionally, reading time can be longer than several minutes in daily scenarios, so gaze drift can significantly affect data quality. In our experiments, we observed that it is difficult for participants to maintain a fixed posture after calibration, though we required them to do so. The posture shift further increases errors. Therefore, in practical applications, real-time calibration of gaze data based on user posture is crucial to ensure data quality. If the existing eye-tracking technology can combined with user posture detection~\cite{faceori}, it is possible to reduce the impact of user posture on gaze data, thereby improving the quality of gaze data.

%% file: sections/8-conclusion.tex
\section{Conclusion}
We propose a highly accurate and real-time unknown-word detecting method based on gaze trajectory and pre-trained language models (PLMs). The text embedding derived from the PLM and the knowledge grounding provides rich linguistic characteristics for candidate words, and the gaze data locates the region of interest and supplies behavior information to the model. The evaluation shows that it achieves an accuracy of 97.6\% and F1-score of 71.1\%. The latency is within 1 second. Our method also works on noisy gaze data acquired by webcam with an accuracy of 97.3\% and an F1-score of 65.1\%, which demonstrates the accessibility and applicability for daily use. The experimental results show that the PLM contributes most to the performance and gaze provides user-dependent features. The real-time evaluation shows that our method achieves the F1-score of 56.54\% and makes reading more fluent. It also shows an improvement in the willingness to use and usefulness compared to the traditional click method. Due to the growing number of embedded eye tracking in commercial devices, we believe that lots of language learners can ultimately benefit from our technique.